\documentclass[11pt,citesort]{article}
\usepackage{epsfig}
\usepackage{amsfonts}
\usepackage{amsmath}
\usepackage{amssymb}
\usepackage{dsfont}
\usepackage{hyperref}
\usepackage{color}
\usepackage{cite}

\newcommand{\al}{\alpha}
\newcommand{\bt}{\beta}
\newcommand{\g}{\gamma}

\newcommand{\simu}{\sigma^{\mu\nu}}
\newcommand{\Fmu}{F_{\mu\nu}}

\newcommand{\slashPT}{\slash\hspace{-0.6em}P\slash\hspace{-0.5em}T}
\newcommand{\slashPTsub}{\slash\hspace{-0.45em}P\slash\hspace{-0.4em}T}

\newcommand{\Nb}{\bar N}
\newcommand{\Fp}{F_\pi}

\newcommand{\tb}{\bar \theta}

\newcommand{\mpi}{m_{\pi}}
\newcommand{\MQCD}{\Lambda_\chi}

\newcommand{\Or}{\mathcal O}

\newcommand{\vL}{\ensuremath{\mathcal{L}}}

\newcommand{\sq}{^{2}}
\newcommand{\ga}{\gamma}
\newcommand{\Lc}{\Lambda_\chi}

\newcommand{\dslash}[1]{#1 \llap{/\kern-0.5pt}}
\newcommand{\Dslash}[1]{#1 \llap{/\kern+1.5pt}}
\newcommand{\DDslash}[1]{#1 \llap{/\kern+2.3pt}}
\newcommand{\dslashh}[1]{#1 \llap{/\kern+1pt}}

\newcommand{\boldtau}{\mbox{\boldmath $\tau$}}
\newcommand{\boldpi}{\mbox{\boldmath $\pi$}}

\newcommand{\CP}{C\hspace{-.5mm}P}
\newcommand{\CPT}{C\hspace{-.5mm}PT}                          %

\newcommand{\bea}{\begin{eqnarray}}
\newcommand{\eea}{\end{eqnarray}}
\newcommand{\bma}{\begin{pmatrix}}
\newcommand{\ema}{\end{pmatrix}}
\newcommand{\nn}{\nonumber}

\newcommand{\efm}{\mbox{$e\,\text{fm}$}}
\newcommand{\ecm}{\mbox{$e\,\text{cm}$}}

\begin{document}
\begin{titlepage}

\vspace{2.0cm}

\begin{center}
{\Large\bf Unraveling models of CP violation through\\[0.3em] electric dipole moments 
of light nuclei}
\vspace{1.7cm}

{\large \bf W. Dekens$^a$, J. de Vries$^b$, J. Bsaisou$^b$, W. Bernreuther$^c$, C. Hanhart$^b$, \\ Ulf-G. Mei{\ss}ner$^{b,d}$, A. Nogga$^b$, A. Wirzba$^b$} 

\vspace{0.5cm}

{\large 
$^a$ 
{\it University of Groningen,
9747 AA Groningen, The Netherlands}}

\vspace{0.25cm}
{\large 
$^b$ 
{\it Institute for Advanced Simulation, Institut f\"ur Kernphysik, 
and J\"ulich Center for Hadron Physics, Forschungszentrum J\"ulich, 
D-52425 J\"ulich, Germany}}

\vspace{0.25cm}
{\large
$^c$ {\it Institut f\"ur Theoretische Physik, RWTH Aachen University, 52056 Aachen,
Germany}}

\vspace{0.25cm}
{\large 
$^d$ 
{\it Helmholtz-Institut f\"ur Strahlen- und Kernphysik and Bethe Center for 
Theoretical Physics, Universit\"at Bonn, D-53115 Bonn, Germany}}

\end{center}

\vspace{1.5cm}

\begin{abstract}
We show that the proposed measurements of the electric dipole moments of light nuclei in storage rings would put strong constraints on
 models of flavor-diagonal CP violation. 
Our analysis is exemplified by a comparison of  the Standard Model including
the QCD theta term, the minimal left-right symmetric model, a specific version of 
  the so-called aligned two-Higgs doublet model, and briefly the minimal supersymmetric extension of the Standard Model. By using effective field theory techniques
we demonstrate to what extent measurements of the electric dipole moments of the nucleons, the deuteron, and helion could
 discriminate between these scenarios. We discuss how measurements of electric dipole moments of other systems relate to the light-nuclear measurements.

\end{abstract}

\vfill
\end{titlepage}

\section{Introduction}
The Standard Model (SM) of particle physics contains in the quark sector two sources of $P$ and $T$ violation\footnote{The models studied in this paper are $\CPT$ invariant. Therefore, $P$ and $T$ violation amounts to $\CP$ violation.}. The best understood source is the phase that is present in the three-generation quark mixing matrix, the Cabibbo-Kobayashi-Maskawa (CKM) \cite{Cabibbo:1963yz,Kobayashi:1973fv} matrix, that induces $\CP$-violating effects in flavor-changing  processes. On the other hand, its contribution to flavor-diagonal $P$- and $T$-odd observables, such as electric dipole moments (EDMs), is highly suppressed and inaccessible with current experimental techniques.
The second $P$- and $T$-violating ($\slashPT$) source is the QCD vacuum angle $\tb$~\cite{'tHooft:1976up, 'tHooft:1976fv} which, in principle, would generate large hadronic EDMs. The null-measurement of the neutron EDM~\cite{Baker:2006ts} strongly limits $\tb \lesssim 10^{-10}$~\cite{Baluni:1978rf, CDVW79}. The puzzle of why $\tb$ is so extremely small or perhaps zero is called the strong $\CP$ problem. In addition, it seems that both $\CP$-odd sources in the SM are unable to account for the current matter-antimatter asymmetry in the universe~\cite{Riotto:1999yt, Kuzmin}. It is therefore believed that the SM cannot be the whole story and that additional $\slashPT$ sources exist. It has been known for a long time that searches for EDMs are highly sensitive probes of additional, flavor-diagonal $\CP$-violating interactions. Excellent reviews on EDMs can be found in Refs.~\cite{Khriplovich:1997ga,Pospelov_review,Engel:2013lsa}.

The above considerations have led to large experimental endeavours to measure EDMs of leptons, hadrons, nuclei, atoms, and molecules (for an overview, see Ref.~\cite{Hewett:2012ns}). At the moment the strongest existing limits have been obtained for the neutron EDM~\cite{Baker:2006ts}, the  EDM of the diamagnetic ${}^{199}$Hg atom~\cite{Griffith:2009zz}, and the electron EDM (inferred from measurements on the polar molecule ThO~\cite{Baron:2013eja}). The main motivations for this work are the plans to measure the EDMs of charged spin-carrying particles in storage rings \cite{Farley:2003wt, Onderwater:2011zz,Pretz:2013us,JEDI}. The spin precession of a particle trapped in such a ring is affected by its EDM and it has been proposed that this method can be used to measure the EDMs of the proton and deuteron with a precision of $10^{-29}\,\ecm$, three orders of magnitude better than the current neutron EDM limit. EDMs of other light ions, such as the helion (${}^3$He nucleus) and triton (${}^3$H nucleus) are candidates as well. 

EDM experiments are very good probes for new $\slashPT$ sources because, as mentioned, at current experimental accuracies they are `background-free' probes of new physics. Any finite signal in one of the upcoming experiments would be due to physics not accounted for by the Kobayashi-Maskawa (KM) mechanism~\cite{Kobayashi:1973fv}. This source of $\CP$ violation induces only very small light quark and nucleon EDMs of the order of $10^{-31}\,\ecm$~\cite{Czarnecki:1997bu,Mannel:2012qk} and even tinier lepton EDMs. A larger EDM signal might be caused by physics beyond the SM (BSM). However, it is not excluded that an extremely small, but nonzero, $\tb$ term could be its origin. An interesting and important problem is therefore to investigate whether it is possible to trace a nonzero $\tb$ with EDM experiments. That is, can we confidently disentangle the $\tb$ term from possible BSM sources\footnote{Solely for the purpose of terminology we distinguish in this paper between the $\tb$ term and BSM sources of $\CP$ violation. Of course, if a small but nonzero $\tb$ exists, it may actually be generated by some BSM dynamics.}?

\begin{figure}[t]
\centering
\includegraphics[scale=0.7]{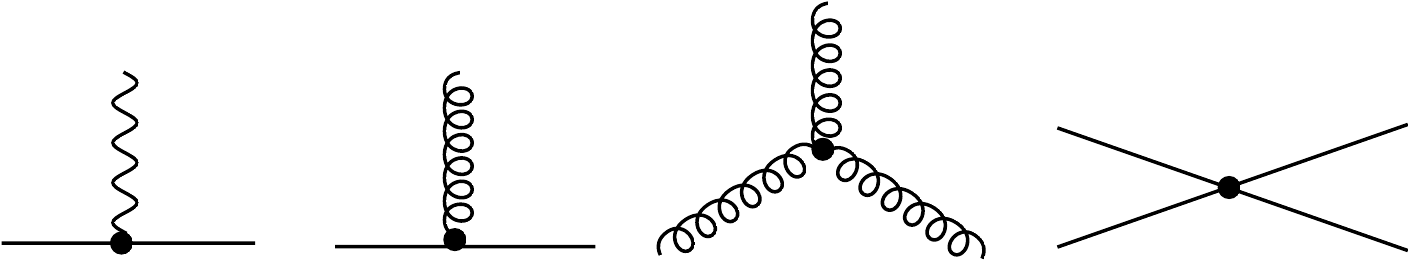} 
\caption{Schematic drawings of the $\slashPT$ dimension-six operators in Eq.~\eqref{introeq}. The QCD $\tb$ term is not shown. The solid, curly, and wavy lines denote external quark, gluon, and photon states respectively. The first and second diagram depict the quark EDM and chromo-EDM, respectively, the third diagram the Weinberg operator, and the fourth diagram a four-quark operator. Vertices with two or more gluon fields associated with the quark CEDM and the Weinberg operator are not shown.} 
\label{introfig}
\end{figure}

To answer this question several obstacles need to be overcome. In order to separate $\tb$ from BSM physics we need a description of the latter. Lacking knowledge of BSM physics, the only model-independent description relies on effective field theory (EFT), which requires the addition of the most general set of $\CP$-violating higher-dimensional operators to the SM Lagrangian. The most important operators are those of dimension six (before electroweak gauge-symmetry breaking)\cite{Buchmuller:1985jz,Grzadkowski:2010es}, while the effects of even higher-dimensional operators are expected to be suppressed. Once the set of effective dimension-six operators has been identified, it needs to be renormalization-group evolved to the low energies where the experiments take place \cite{Wilczek:1976ry, BraatenPRL, Degrassi:2005zd, Hisano3, Dekens:2013zca}. The evolution of the effective operators can be calculated in perturbation theory only  down to a scale $\Lambda_\chi$ of the order of $1$ GeV. Below this scale, the expansion in the strong coupling constant breaks down and nonperturbative techniques become necessary. 
At the scale $\Lambda_\chi$, the $\slashPT$
low-energy effective Lagrangian of the quark and gluon
  degrees of freedom schematically takes on the form (see also Fig.~\ref{introfig}):
\begin{eqnarray}\label{introeq}
\mathcal L_{\slashPTsub} &=& - \tb \frac{g^2}{64\pi^2}\epsilon^{\mu\nu\alpha\beta} G^a_{\mu \nu}G^a_{\alpha \beta}  -\frac{1}{2} \sum_{q={u,d}}\left( d_q \, \bar q i\simu \ga_5 q\, F_{\mu\nu} + \, \tilde d_q\, \bar q i\simu \ga_5 t_a q\, G^a_{\mu\nu}\right)\nonumber\\
&&+ \frac{d_W}{6} f_{abc}\varepsilon^{\mu\nu\al\bt}G^a_{\al\bt}G^b_{\mu\rho}G_{\nu}^{c \, \rho} + \sum_{i,j,k,l = {u,d}}C_{ijkl}\, \bar q_i \Gamma  q_j \, \bar q_k \Gamma^\prime  q_l\,\,\,,
\end{eqnarray} 
in terms of the quark fields $q$, the photon and gluon field-strength
tensors $F_{\mu\nu}$ and $G_{\mu\nu}^a$, respectively. The
  $f_{abc}$ are the structure constants and the $t_a$ are the
  generators in the fundamental representation of $SU(3)_c$. The coefficients $d_q$ and $\tilde d_q$ are the
  electric dipole  and chromo-electric dipole moments (CEDM) of
  quarks, and the coefficient $d_W$ of the Weinberg operator
  \cite{Weinberg:1989dx} can be interpreted
  as the chromo-electric dipole moment
  (gCEDM) of the gluon \cite{BraatenPRL,Braaten:1990zt}.
 The last term contains four-quark operators with zero net-flavor
 where the  matrices $\Gamma$ and $\Gamma^\prime$ denote various Lorentz structures such that the four-quark operators violate the $\CP$ symmetry. In this work, we consider the low-energy $\slashPT$ Lagrangian for $u$ and $d$ valence quarks only, which is appropriate for analyzing the EDMs of nucleons and light nuclei. The first operator in Eq.~\eqref{introeq} is the dimension-four QCD $\tb$ term, while the others are or arise from dimension-six operators (before electroweak gauge-symmetry breaking) 
and are generated by BSM dynamics. The second and third operators are the quark EDMs and chromo-EDMs respectively, the fourth operator is the Weinberg operator~\cite{Weinberg:1989dx}, and the last term denotes various $\slashPT$ four-quark operators.  We will discuss these operators in much more detail in the subsequent sections. 

Below the scale $\Lambda_\chi$, EFTs are again a very powerful tool in
understanding low-energy strong interactions. By constructing the most
general interactions for the low-energy degrees of freedom which are
consistent with the symmetries of QCD, chiral symmetry in particular,
and with their spontaneous and explicit breakdown, it is possible to obtain an effective low-energy description of QCD called chiral perturbation theory ($\chi$PT) \cite{Weinberg:1978kz,Gasser:1983yg,Gasser:1987rb, Weinbergbook, BKM95}. The main advantage of $\chi$PT is that observables can be calculated perturbatively with an expansion parameter $q/\Lambda_\chi$ where $q$ is the typical momentum scale of the process under consideration.  
Each interaction appearing in the chiral Lagrangian is associated with a low-energy constant (LEC) whose size is not fixed by symmetry considerations and depends on the strong nonperturbative dynamics. However, the perturbative nature of $\chi$PT ensures that most observables only depend on a small number of LECs.  Once these LECs have been determined, either by fitting them to data or by direct lattice calculations, other observables can be firmly predicted. 
Another major success of $\chi$PT is the description of the nucleon-nucleon and multi-nucleon interactions. This has opened up the way to describe nucleons and (light) nuclei in a unified framework \cite{Bedaque:2002mn, Epelbaum:2008ga, Machleidt:2011zz}.  

In recent years, $\chi$PT has been extended to include effects of the $\tb$ term~\cite{BiraEmanuele} and 
 $\slashPT$  BSM operators up to  dimension-six~\cite{deVries:2012ab}  which induce $\slashPT$ interactions in the chiral Lagrangian. 
The so-amended $\chi$PT allows for the calculation of the EDMs of the nucleon~\cite{Pich:1991fq, Borasoy:2000pq,BiraHockings, Narison:2008jp,ottnad,deVries2010a, Mer11,  Guo12} and light nuclei~\cite{deVries2011a, deVries2011b, Jan_2013} in terms of the various LECs associated with the $\slashPT$ chiral interactions. The nuclear uncertainties can be quantified and improved upon systematically. Although the hadronic uncertainty in  the sizes of the LECs themselves is sizable, the same LECs appear in several EDMs which means that the hadronic uncertainties cancel to a large degree.  It is this property, in addition to the high envisaged experimental accuracy, which makes the plans to measure the EDMs of light nuclei so exciting. Calculations of EDMs of heavier systems, such as ${}^{199}$Hg, suffer from much larger nuclear uncertainties which are hard to quantify~\cite{deJesus:2005nb,Engel:2013lsa}.

Although the $\tb$ term and the higher-dimensional operators in Eq.~\eqref{introeq}
 all break $P$ and $T$, they transform differently under chiral and isospin rotations. This ensures that the 
different \mbox{$\slashPT$}  sources induce different $\slashPT$ chiral Lagrangians, which, in turn, lead to distinct patterns of EDMs.  
This observation has been used in recent works which concluded that it is possible to disentangle the $\tb$ term from the higher-dimensional BSM operators,
given enough EDM measurements \cite{Pospelov_deuteron,deVries2011a, deVries2011b, Jan_2013 }. In particular, the EDM of the deuteron plays an important role. Furthermore, several 
classes of dimension-six operators can be disentangled among themselves as well. Again the deuteron EDM plays an important role, 
 but the EDMs of the helion and/or triton give  important complementary information. 

The investigation of light-nuclear EDMs so far has focused mainly on
the $\tb$ term and dimension-six operators individually. That is, it
was assumed \cite{deVries2011a,deVries2011b,Jan_2013,deVries:2012ab}
that one operator is dominant over the others which has the advantage
of a rather clean analysis. It can be questioned, however, how realistic such a scenario is. It could very well be that the underlying 
microscopic theory induces contributions of similar size to several effective dimension-six operators. Furthermore, even if only one operator 
turns out to be dominant at high energies, this operator can  induce sizable contributions to other operators when evolved to the low-energy scale where 
EDM experiments take place. Therefore, the assumption of one dominant dimension-six operator at low energies might not be the most likely one. 
To investigate this in more detail, we study in this work four distinct scenarios of non-KM $\CP$ violation and investigate whether EDM 
measurements can discriminate between them. However, the methods used are in no way limited to these four scenarios and 
 can be easily applied to other BSM models. 

In the first scenario we assume the SM $\tb$ term to be the dominant source. Since, with
the exception of the CKM matrix, this is the only
$\CP$-violating term of dimension four in the hadronic sector, it
provides the background to which the other scenarios, which induce
$\slashPT$ operators with dimension of
     at least of dimension six, have to be compared to. The $\tb$ scenario has already been studied extensively in the 
literature (although we will consider here some very recent results on light-nuclear EDMs~\cite{J.Bsaisou, Jan_new}) and we will
 mainly summarize the results in the following. For the BSM models discussed in this paper, we assume that the $\tb$ term is absent, 
for instance, due to a Peccei-Quinn symmetry~\cite{Peccei:1977hh,Peccei:1977ur} of the Lagrangian of the respective model.

The second scenario is the minimal left-right symmetric
model~\cite{LRSM,Mohapatra:1974hk,Zhang:2007da}. In this model, parity
is restored at energies above the electroweak scale by extending the
SM gauge symmetry to include $SU(2)_R$. It turns out that in this
model the dominant contribution to the respective 
 $\slashPT$ Lagrangian at high energies is due to one particular dimension-six operator. 
This operator mixes with only one additional operator such that the low-energy Lagrangian 
at the quark level is rather simple. However, these operators transform in a rather complicated way under 
chiral symmetry. As a result, the induced chiral Lagrangian contains some interesting and nontrivial structures. 
These structures induce a profound hierarchy of nuclear EDMs which is quite distinct from the $\tb$ scenario. 

The third scenario we investigate is based on the so-called aligned two-Higgs-doublet model (a2HDM)~\cite{Pich:2009sp}. 
Contrary to the two scenarios just outlined, in this model, which exemplifies the generic feature of non-KM $\CP$ violation in two-Higgs doublet-models, 
several $\slashPT$ operators are induced at the level of quarks and gluons which, in general, make contributions of comparable size to hadronic EDMs. 
The coefficients of these operators depend on different parameters of the model.
 The main goal of this work to show that the EDMs of nucleons and light nuclei can be used to disentangle different scenarios, and we do not aim at a 
fully detailed analysis of the a2HDM. We therefore make certain assumptions \cite{Jung:2013hka} regarding the neutral Higgs sector such that all 
 induced higher-dimensional BSM operators depend on the same combination of parameters. 
Despite this simplification, the EDMs of nucleons and light nuclei receive comparable contributions from three BSM  operators which 
makes the analysis more complicated and uncertain. Nevertheless, we demonstrate that the model still leads to a 
 different hierarchy of EDMs than the previous scenarios. 

 Furthermore, we shortly discuss another popular BSM model with non-KM $\CP$ violation, the minimal supersymmetric extension  (MSSM) of the Standard Model. Also in this model, the contribution to the 
 hadronic EDMs is, in general, not dominated by just one $\slashPT$ operator at the level of quarks and gluons.

We will show that estimates of the nucleon EDMs alone are insufficient to disentangle these scenarios. In fact, the 
 predictions and estimates of the two-nucleon contribution to the EDMs of the  light ions,  especially of the 
 deuteron and helion, will be crucial in disentangling the various sources. The measurements
of the deuteron and helion EDM provide in this regard `orthogonal' information, because the
 isospin-filter property of the deuteron favors isospin-breaking interactions, while the helion allows
for both isospin-conserving and -breaking contributions. 
 
This article is organized as follows. In Sect.~\ref{3scenarios} we
discuss  the four different scenarios of $\CP$ violation outlined
above. In particular we focus on
the low-energy $\slashPT$ interactions that are induced in these
scenarios. In Sect.~\ref{chiral} we discuss the most important
$\slashPT$ hadronic interactions that appear in each of the
scenarios. In particular we focus on the $\slashPT$ pion-nucleon
interactions and the nucleon EDMs. In Sect.~\ref{lightnuclei} we turn
to the EDMs of light nuclei. We argue that chiral effective field
theory is a powerful tool to study these observables and show that
measurements of light-nuclear EDMs can be used to disentangle
different scenarios. In Sect.~\ref{sec:other} we 
 briefly discuss other systems, in particular the EDMs of the electron
 and the diamagnetic atom ${}^{199}$Hg. 
 We summarize, conclude, and give an outlook in Sect.~\ref{disc}. 
 Several appendices are devoted to technical details.

\section{Four scenarios of $\CP$ violation}
\label{3scenarios}
In this section we discuss in detail  four distinct scenarios of $\CP$ violation. In particular we discuss the low-energy $\slashPT$ operators that are induced in each scenario. In this work we mainly focus on the EDMs of nucleons and light nuclei. Therefore we concentrate here on the $\slashPT$ operators involving quark and/or gluon fields, while (semi-)leptonic operators are discussed in Sect.~\ref{sec:other}. 
\subsection{The QCD $\tb$ term}\label{theta}
The QCD Lagrangian for two quark flavors is given by
\begin{equation}\label{QCD1}
\mathcal L_{\mathrm{QCD}} = -\frac{1}{4}G_{\mu\nu}^aG^{a,\mu\nu} + \bar q(i\Dslash{D} - M)q - \tb \frac{g^2}{64\pi^2}\epsilon^{\mu\nu\alpha\beta} G^a_{\mu \nu}G^a_{\alpha \beta}\,\,\,,
\end{equation}
 where  $q = (u\, ,  d)^T$  denotes the quark doublet of up
   and down quarks. As already mentioned above, the restriction to two
   quark flavors is appropriate for analyzing the EDMs of nucleons and
   light nuclei within the framework of $\chi$PT.
 In Eq.~\eqref{QCD1}   $G^a_{\mu\nu}$ is the gluon field strength tensor, $\epsilon^{\mu\nu\alpha\beta}$ ($\epsilon^{0123}=+1$)
 is the completely antisymmetric tensor in four dimensions, $D_\mu$ the gauge-covariant derivative, $M$ the 
real-valued quark $2\times 2$ mass matrix, and $\tb$ the coupling
constant of the so-called $\tb$ term which violates $P$ and $T$. 
 In this expression, we have absorbed the complex phase of the quark mass matrix into $\tb = \theta + \mathrm{arg}\,\mathrm{det}(M)$.
Due to the $U_A(1)$ anomaly, an axial $U(1)$ transformation of the
quark fields can be used to remove the $\tb$ term from the
Lagrangian. 
 After vacuum alignment \cite{Baluni:1978rf} and assuming $\tb \ll 1$, the QCD Lagrangian becomes
\begin{equation}\label{QCD2}
\mathcal L_{QCD} = -\frac{1}{4}G_{\mu\nu}^aG^{a,\mu\nu} + \bar q i\Dslash{D}  q -\bar m \bar q q - \varepsilon \bar m \bar q \tau_3 q + m_\ast \tb \bar q i\gamma^5 q\,\,\,,
\end{equation} 
in terms of the averaged quark mass  $\bar m = (m_u+m_d)/2$ in the two-flavor case, 
the quark-mass difference 
$\varepsilon = (m_u-m_d)/(m_u + m_d)$, and the reduced quark mass
$m_\ast = m_u m_d/(m_u + m_d) = \bar m (1-\varepsilon^2)/2$. 
This expression shows that $\slashPT$ effects due to the $\tb$
 term would vanish if one of the quarks were massless.
  However, this is not realized in nature~\cite{pdg:2012}. 
We also give the explicit $PT$-even quark mass terms here because, as we will discuss later, 
\mbox{$\slashPT$} hadronic interactions induced by the $\tb$ term are closely linked to $PT$-even isospin-breaking interactions induced by the quark mass difference. 

Before continuing the analysis of the $\tb$ term, we first discuss the
BSM scenarios used in this paper.
  In these scenarios we assume that the $\tb$ term is absent, for
  example, due to the Peccei-Quinn
  mechanism~\cite{Peccei:1977hh,Peccei:1977ur} or by extreme
  fine-tuning. It should be noted that the Peccei-Quinn mechanism
  would, apart from removing the $\tb$ term, also affect the
  dimension-six operators appearing in the other scenarios
  \cite{Pospelov_review}. 

\subsection{The minimal left-right symmetric model}\label{mLRSM}
Left-right symmetric (LR) models are based on the gauge group
$SU(3)_c\times SU(2)_L \times SU(2)_R \times U(1)_{B-L}$ with an
unbroken parity symmetry at 
high energies \cite{LRSM,Mohapatra:1974hk,Senjanovic:1975rk,Senjanovic:1978ev,
  Deshpande:1990ip}. The abelian subgroup is associated with baryon
minus lepton number. The left-handed and right-handed quarks and
leptons form fundamental representations of $SU(2)_L$ and $SU(2)_R$, respectively.
 As a consequence, right-handed neutrinos are introduced automatically:
\bea Q_L &=&\bma U_L\\D_L\ema \in (3,2,1,1/3)\,\,, \qquad Q_R = \bma U_R\\D_R\ema\in (3,1,2,1/3)\,\,,\nn\\
L_L &=&\bma \nu_L\\l_L\ema \in (1,2,1,-1)\,\,, \qquad L_R = \bma \nu_R\\l_R\ema\in (1,1,2,-1)\,\,\,,\eea
where the capital letters $Q$, $U$,  $D$, and $L$
  denote quarks or leptons of any generation.
Given the fermion assignment, at least one spin-zero bidoublet,
$\phi$, with the assignment $(1,2,2,0)$, is needed to generate
fermion masses.  The LR model is called \textit{minimal}
  \cite{Zhang:2007da} if just one bidoublet is used such that the
  model is parity-invariant before gauge-symmetry breaking, but $\CP$ is
  broken both explicitely and spontaneously.

At some high-energy scale above the electroweak scale
 the extended gauge-group of the LR model should be broken down to the
 SM gauge-group. In order to achieve this, additional spin-zero fields
 are employed. 
In the version of the minimal LR model (mLRSM) we will discuss here,
{\it cf.} \cite{Deshpande:1990ip, Zhang:2007da}, 
this is done with two triplets $\Delta_{L,R}$ belonging to $(1,3,1,2)$ and $(1,1,3,2)$,
 respectively. The spin-zero fields can be written in the form
\bea \phi = \bma \phi_1^0 & \phi_2^+\\ \phi_1^- & \phi_2^0 \ema ,\qquad
\Delta_{L,R} = \bma \delta^+_{L,R}/\sqrt{2} & \delta^{++}_{L,R}  \label{LRphi}\\ 
\delta^0_{L,R} & -\delta^+_{L,R}/\sqrt{2} \ema\,\,\,. \label{LRDelta}
\eea
With this definition of the fields the parity transformation is
equivalent  to changing the (L, R) indices of all fields to (R, L) and letting
$\phi \rightarrow \phi^\dagger$. 
Among other things this symmetry implies that the coupling constants
of the two $SU(2)$ gauge-groups are equal.

In order to achieve the breaking of both the gauge symmetries and
  the parity symmetry, the neutral  components of the  spin-zero fields
  are assumed to acquire vacuum expectation values (vevs). First, the
  symmetry group $SU(3)_c\times SU(2)_L \times SU(2)_R \times
  U(1)_{B-L}$ is broken to  down to the SM gauge group,
  $SU(3)_c\times SU(2)_L \times U(1)_{Y}$ by the vev $\langle
  \Delta_R\rangle =v_R$ at a scale of several TeV. This gauge-symmetry
  breaking entails also the breaking of the parity symmetry. 
 The vev $v_R$ sets the scale of the masses of the additional gauge bosons, $W_{R}^\pm$ and $Z_R$, of the $SU(2)_R$ gauge group. 
 In order for the mLRSM to satisfy the experimental bounds coming from $K$- and $B$-meson mixing,  the mass of the right-handed $W_R^\pm$ boson is constrained to $M_{W_R}\geq 3.1 \, \text{TeV}$ \cite{Bertolini:2014sua}.
 At lower energies, electroweak symmetry breaking is achieved by the
 vevs of the bidoublet $\phi$.
 Lastly, the vev $\langle\Delta_L\rangle=v_L$ gives rise to a Majorana mass term for
the left-handed neutrinos. This implies that this vev should not be much
larger than the scale of the neutrino mass, $v_L \lesssim
\Or(\text{eV})$. The vev $v_L$ and its phase $\theta_L$, however, do
not enter in the terms in the Lagrangian in Eq.~\eqref{introeq}  which are important for
hadronic EDMs. Therefore, they will not play a role in any of our
calculations below. Explicitly, the spin-zero fields acquire the
following vacuum expectation values  with two observable $\CP$-violating
phases, which by convention, are put into the vev of the second
doublet and of $\Delta_L$  \cite{Deshpande:1990ip,Zhang:2007da}:
\bea
\langle \phi \rangle = \bma \kappa &0\\0&\kappa' e^{i\al}
\ema\,\,,\qquad \langle \Delta_{L} \rangle = \bma
0&0\\v_{L}e^{i\theta_L}&0\ema\,\,,\qquad \langle \Delta_{R} \rangle =
\bma 0&0\\v_{R}&0\ema\,\,\, .
\label{vev}\eea
The vevs  $\kappa, \kappa'$ set the scale of the masses of the $W_L^\pm$ and $Z_L$ gauge
bosons of  the $SU(2)_L$ gauge group. We have
\bea\sqrt{2}\sqrt{\kappa\sq +\kappa^{\prime 2}} = v \simeq 246 \, \text{GeV}\,\,\,.\eea 

$\CP$ violation in the quark sector of the mLRSM arises from a
number of phases.  From explicit and spontaneous $\CP$ breaking in the
 Higgs potential, the $\CP$-violating phase  $\al$ of Eq.~\eqref{vev}
 is generated \cite{Deshpande:1990ip}.
  Secondly, additional phases will
appear in the  quark mixing matrices $V_L$ and $V_R$. 
The matrix  $V_L$ of the left-handed quarks, which is identical
to the CKM matrix of the SM, contains one observable phase. 
 Similarly, a right-handed
analogue of the CKM matrix is produced when the quark mass-eigenstates
are not aligned with the $SU(2)_R$ eigenstates, which  will
be the case in general. In addition to the KM  phase in $V_L$,
there are then six additional phases in $V_R$.
 However, in order to produce the correct pattern 
 of quark masses the model parameters have to be tuned in such a way
 that there is an approximate relation between the two quark mixing
  matrices and their phases \cite{Maiezza:2010ic}.

Having discussed the model we are now ready to integrate out the heavy
fields and derive the dimension-six $\CP$-odd operators produced at
the electroweak scale that are relevant for the effective Lagrangian
in Eq.~\eqref{introeq}.
 The phases in $V_L$ and $V_R$,
together with the phase $\al$, produce a number of $\CP$-violating
operators at the electroweak scale. However, just one of these
  operators is generated at tree level, while the others are
    induced at the one-loop level. Hadronic EDMs in the mLRSM are
  therefore dominated by this single operator. 
  For a more detailed discussion we refer to Appendix~\ref{LRmodel}
  and Refs.~\cite{Xu:2009nt,An:2009zh}. Below the scale of the new
  physics, this dominant 
interaction takes the form of the following gauge-invariant Lagrangian \cite{Grzadkowski:2010es, Ng:2011ui},
\bea
\vL_{LR}\big|_{\mu \lesssim v_R} = \Xi_1 (i\tilde\varphi^{\dagger}D_\mu \varphi)(\overline u_R \g^\mu d_R) +\text{h.c.} \,\,,\qquad
\Xi_1 = \frac{2}{v\sq}\frac{\kappa \kappa'}{v_R\sq} V_R^{ud }e^{i\al}\,\simeq -\frac{2}{v\sq}\sin\zeta\, V_R^{ud }e^{i\al}\,\,\,,\label{Wcurrent}
\eea
where $\varphi$ 
corresponds to  the SM Higgs-doublet (see Appendix~\ref{LRmodel}), $\tilde\varphi=i\tau_2\varphi^*$,
 and $\mu\lesssim v_R$ indicates the scale where the above effective
 Lagrangian describes the dominant $\CP$ violation in this model. Furthermore, $\sin\zeta \simeq -\kappa \kappa'/v_R\sq$ is the angle describing the mixing between the $W_L^\pm$ and $W_R^\pm$ bosons, see, for instance, Ref.~\cite{Zhang:2007da}. After
 electroweak symmetry breaking, this operator becomes in the unitary gauge
\bea\label{WcurrentEW}
\vL_{LR}\big|_{\mu \sim v} =-\frac{gv\sq}{2\sqrt{2}}\bigg[ \Xi_1\,
\overline u_R \g^\mu d_R W_{L\mu}^+ +
\text{h.c.}\bigg]\bigg(1+\frac{h}{v}\bigg)^2\,\,\, ,
\eea
where $h$ is the lightest Higgs boson of the model, i.e., it corresponds to the 126 GeV
 spin-zero resonance discovered at the LHC \cite{Aad:2012tfa,Chatrchyan:2012ufa}.

The above interaction is essentially a coupling of the $W_L^\pm$ boson
to right-handed quarks. 
This interaction is generated because both $W_L^\pm$ and $W_R^\pm$
interact with the bidoublet $\phi$, as this field 
is charged under $SU(2)_L$ and $SU(2)_R$. Through their interactions
with the bidoublet, $W_L^\pm$ and $W_R^\pm$ effectively 
mix among each other. Thus, after integrating out the heavy $W_R^\pm$
boson, the remaining $W_L^\pm$ boson obtains a (small) 
coupling to right-handed fields in the form of the operator in Eq.~\eqref{WcurrentEW}. 

The operator in Eq.\ \eqref{Wcurrent} remains invariant under
  QCD renormalization-group evolution. Therefore we can trivially
lower the energy to the electroweak scale. In order 
to move to even lower energies, we need to integrate out the heavy SM
fields. 
 Integrating out the $W_L^\pm$ and Higgs fields, we obtain, just below the mass of the $W_L^\pm$ boson, 
\bea
\vL_{LR}\big|_{\mu\lesssim m_W} = - i\,\text{Im}\left(V_L^{ud*}\,\Xi_1(M_W) \right)\, \left(\overline u_R \g^\mu d_R\,\overline d_L \g_\mu u_L -\overline d_R \g^\mu u_R\,\overline u_L \g_\mu d_L\right)+\dots\,\,\,,
\eea
where the dots denote suppressed operators. The resulting four-quark
operator is affected by QCD corrections and, in fact, mixes with a
second operator which has the same Lorentz but  different color structure.
At a scale where perturbative QCD is still valid, well above the chiral scale $\Lambda_\chi \sim 1\,\text{GeV}$,
we obtain,
\bea\label{mLRSM1GeV}
\vL_{LR}\big|_{\mu\sim \Lambda_\chi} &=&   -i \eta_1 \,\mathrm{Im}\,\left(V_L^{ud*}\,\Xi_1(M_W)\right)\left(\overline u_R \g^\mu d_R\,\overline d_L \g_\mu u_L -\overline d_R \g^\mu u_R\,\overline u_L \g_\mu d_L\right)\\
&&-i \eta_8 \, \mathrm{Im}\,\left(V_L^{ud*}\,\Xi_1(M_W)\right)\left(\bar u_{R}\ga^\mu t_a d_{R}\, \bar d_{L}\ga_\mu t_a u_{L}-
\bar d_{R}\ga^\mu t_a u_{R}\, \bar u_{L}\ga_\mu t_a d_{L}\right)+\dots\,\,\,,\nn
\eea
where  $\eta_1 = 1.1$ and
$\eta_8 =1.4$ are factors appearing due to QCD evolution
\cite{Dekens:2013zca}. As these four-quark operators contribute to
hadronic EDMs, their coefficients can be bounded by the upper limit on
the neutron EDM. This gives
$v\sq\mathrm{Im}\,\left(V_L^{ud*}\,\Xi_1(M_W)\right)\leq 8 \cdot
10^{-5}$, see Sect.~\ref{sec:bounds}. A stronger bound was found in Ref.~\cite{Zhang:2007da},
but a a recent $\chi$PT analysis indicated that the strength of this upper
bound has been overestimated \cite{Seng14}. 
In any case, in the mLRSM,  the dominant
$\CP$-violating contribution
to the effective Lagrangian in Eq.~\eqref{introeq}
 at the chiral scale is given  by the combination of four-quark operators in Eq.~\eqref{mLRSM1GeV}.

\subsection{The aligned two-Higgs-doublet model}
 \label{2HDM}
 Two-Higgs-doublet models (2HDMs) are among the simplest
   extensions of the SM. Among other features they provide an
   interesting source for non-KM $\CP$ violation, namely $\CP$ violation
   induced by neutral and charged Higgs boson exchange, for reviews,
   see, e.g., Refs.~\cite{Gunion:1989we,Bigi:2000yz,Branco:2011iw}.
 In these models the  SM field content
 is extended by an additional Higgs doublet. There are thus two
 doublets, 
$\Phi_1$ and $\Phi_2$, both transforming under $SU(3)_c\times SU(2)_L\times U(1)_Y$ as $(1,2,1/2)$.
{The electroweak symmetry is broken by the vevs of the neutral
  components of $\Phi_1$ and $\Phi_2$.
   One can always choose a so-called Higgs
basis (see, for instance, Ref.~\cite{Branco:2011iw}),} in which only one of the doublets acquires a vev,
\bea \langle \Phi_1 \rangle =  (0,v/\sqrt{2})^T\,\,,\qquad \langle \Phi_2 \rangle = 0\,\,\,,\eea
where $v=\sqrt{v_1\sq +v_2\sq}\simeq 246 \, \text{GeV}$. 
 In this basis
the {would-be Goldstone boson fields $G^0$ and $G^+$ are contained in $\Phi_1$:}
\bea 
\Phi_1 = \bma G^+\\ (v+S_1+i G^0)/\sqrt{2}\ema\,\, ,\qquad
\Phi_2 = \bma H^+\\ (S_2+i S_3)/\sqrt{2}  \ema\,\,\,.
\eea
Thus, the physical spin-zero fields of the 2HDM consist of $3$ neutral
fields, $S_{1,2,3}$, and one charged field, $H^+$. 
{The neutral fields in the mass basis, $\varphi^0_1, \varphi^0_2,
  \varphi^0_3$, are   linear combinations of the fields
  $S_i$. 
  The two sets of fields are related by an orthogonal $3\times 3$ matrix
 $R$, $\varphi_i^0 =R_{ij}S_j$.
 In general, the Higgs potential of a
  2HDM  violates $\CP$. As a consequence, the $\varphi^0_i$ do not have a definite $\CP$
  parity. The lightest of the fields $\varphi^0_i$ 
  corresponds to the 126 GeV spin-zero resonance discovered at the LHC
  \cite{Aad:2012tfa,Chatrchyan:2012ufa}. 
 If the Higgs potential conserves $\CP$, then two of the  $\varphi^0_i$  have $\CP$ parity
$+1$, while the third one has  $\CP$ parity $-1$. In this case, the
fields in the mass basis are traditionally denoted by $h,\,H,\,A$,
where $h$ describes the lightest of the two scalar states. 
 For the sake of simplicity, we shall use  here the notation
 $(\varphi^0_i)=(h,\,H,\,A)$  also in the case of
 Higgs sector $\CP$ violation, when $h$, $H$, and $A$ no longer have
 a definite $\CP$ parity.}

The most important contributions to EDMs arise from the interactions
of  the spin-zero fields with fermions. {The most general Yukawa
  Lagrangian in the quark sector that obeys  the SM gauge symmetries is given by
\bea 
-\vL_{Y'} = \frac{\sqrt{2}}{v}\left[\bar Q'_L (M'_d \Phi_1 +Y'_d\Phi_2)D'_R +
  \bar Q'_L (M'_u \tilde\Phi_1 +Y'_u\tilde\Phi_2)U'_R
  +\text{h.c.}\right]\,\,\, . \label{yukawa2}
\eea
 Here $Q'_L$, $U'_R$, and $D'_R$ denote the $SU(2)_L$ quark
  doublet and singlets, respectively, in the weak-interaction basis,
 $M'_{u,d}$ and $Y'_{u,d}$ are complex $3\times 3$ matrices;
$M'_{u,d}$ are the quark mass matrices to which $Y'_{u,d}$, in view of
having chosen the Higgs basis,  do not contribute.}

So far we have discussed a general 2HDM. {The requirement for
restricting these models comes from the fact that 
in its general form the 2HDM generates tree-level flavor-changing
neutral-currents \mbox{(FCNCs)}~\cite{Gunion:1989we,Bigi:2000yz,Branco:2011iw}. 
 One way to make sure these tree-level FCNCs are absent is to impose
 a  $Z_2$ symmetry on the model which may be softly broken by the
 Higgs potential~\cite{Gunion:1989we,Bigi:2000yz,Branco:2011iw}.
 An alternative and  more general way is
to assume that the matrices $M'_{q}$ and $Y'_{q}$ $(q=u,d)$ are, at some
 (high) scale, 
proportional to each other and can therefore be simultaneously diagonalized \cite{Pich:2009sp}:}
\bea \label{eq:speca2hdm}
Y_d = \varsigma_d M_d\,\,,\qquad Y_u = \varsigma_u^* M_u\,\,\,,
\eea
where $M_{u,d}$ are the
  real diagonal quark mass matrices and $\varsigma_u$ and $\varsigma_d$ are complex numbers. 
  The model using this
 assumption is called the aligned two-Higgs doublet model (a2HDM) \cite{Pich:2009sp}. This is similar to the hypothesis 
of `minimal flavor violation'~\cite{D'Ambrosio:2002ex, Buras:2000dm,Buras:2010zm}.

In the mass basis both for the quark and the Higgs fields, one obtains
from Eq.~\eqref{yukawa2}, using  Eq.~\eqref{eq:speca2hdm}, the Yukawa
interactions of the quark and the physical Higgs fields $H^\pm$ and $\varphi^0_i$:
\bea
-\vL_{Y}& =& \frac{\sqrt{2}}{v}H^+\bar U \big[\varsigma_d V {M}_d \mathcal{P}_R - \varsigma_u {M}_u V \mathcal{P}_L\big]D\nn\\
&+ &\frac{1}{v}\sum\limits_{i=1}^3 \bigg[y_u^i\varphi_i^0\, \bar U_L
{M}_u U_R+y_d^i\varphi_i^0\, \bar D_L {M}_d D_R\bigg]
+\text{h.c.} \quad ,
\label{yukawa3}
\eea
where $V$ is the CKM matrix and $\mathcal{P}_{R,L} = (1\pm\g_5)/2$. The reduced Yukawa couplings
  $y_u^i$ and $y_d^i$ of the neutral Higgs bosons are given in terms
  of the complex parameters $\varsigma_{u,d}$ and the  matrix
  elements of the $3\times 3$ real orthogonal Higgs mixing matrix $R$ (see above)
  by   \cite{Pich:2009sp}:
\bea 
y_u^i &=&R_{i1} +(R_{i2}-i R_{i3})\varsigma_{u}^* \,\, , \label{eq:y0u2hdm}
\\
y_d^i &=& R_{i1} +(R_{i2}+i R_{i3})\varsigma_{d} \,\, . \label{eq:y0d2hdm}
\eea
The orthogonality of $R$ implies
\bea
\sum\limits_{i=1}^3  \text{Re}\, y_q^i\,\text{Im}\, y_{q'}^i = r_{q'} \text{Im}\, (\varsigma_{q}^*\varsigma_{q'})\,\,\,,
\label{simplification}
\eea
where $r_u=-1$, $r_d=1$.
The Yukawa couplings to leptons are analogous to those of the down-type
  quarks. Resulting $\CP$-violating effects involving leptons effects will be commented on in 
Sect.~\ref{sec:eEDM} and in Appendix~\ref{A2HDMAppendix}.

The interactions in Eq.~\eqref{yukawa3} and the couplings  in Eqs.~\eqref{eq:y0u2hdm} and \eqref{eq:y0d2hdm}
exhibit several interesting features.  a) The exchange of a charged
Higgs boson between quarks transports, apart from the KM-phase that plays
  no role in the discussion below, also
the $\CP$-violating phases of $\varsigma_{u,d}$. These phases induce, for
instance, flavor-diagonal $\CP$-odd four-quark operators already at tree-level
 of the type $(\bar{u}d)(\bar{d}i\gamma_5 u)$, where $u$ $(d)$
   denotes any of the up-type (down-type) quarks, with operator
   coefficients proportional to ${\rm
     Im}(\varsigma_{u}^*\varsigma_{d})$. b) If the Higgs potential
   violates $\CP$, the neutral Higgs states are, as mentioned, no longer
   $\CP$ eigenstates and their exchange induces, for
 instance,   flavor-diagonal $\CP$-odd four-quark operators
 $(\bar{q}q) (\bar{q}'i\gamma_5 q')$, $(q, q' =u, d)$ at tree level, 
  in particular the operators
 $(\bar{q}q)(\bar{q}i\gamma_5 q)$. c) If the (tree level) Higgs
 potential of the a2HDM
 is $\CP$-invariant, neutral Higgs exchange  nevertheless induces 
  $\CP$-odd operators of the  type  $(\bar{u}u) (\bar{d}i\gamma_5 d)$
  if  ${\rm Im}(\varsigma_{u}^*\varsigma_{d})\neq 0.$
 Features a) and c) distinguish the a2HDM from 2HDM with a $Z_2$
 symmetry that is (softly) broken by $\CP$-violating Higgs potential, \textit{cf}., for
 instance Ref.~\cite{Bernreuther:1992dz,Inoue:2014nva}.
     
\begin{figure}[t]
\centering
\includegraphics[scale=0.8]{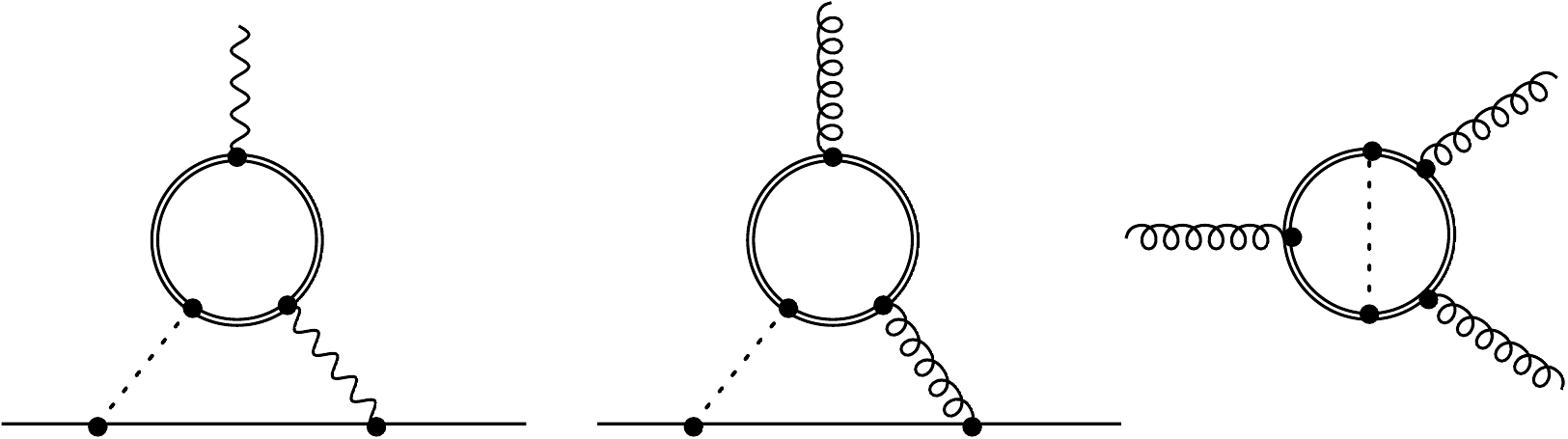} 
\caption{
    Examples of two-loop diagrams {which contribute to the
    coefficients of the operators in Eq.~\eqref{2HDM1GeV}.
   A single (double) solid  lines denotes
   a light (heavy) quark, a dashed line corresponds to a Higgs
   boson,  and a wavy and a curly line depicts a  photon and a  gluon,
    respectively. Diagrams (a,b) are Barr-Zee type diagrams
    contributing to the quark electric and chromo-electric dipole
    moment, while diagram (c) contributes to the Weinberg 
    operator.}
} 
\label{diagrams}
\end{figure}

\begin{figure}[t]
  \centering
 \includegraphics[width=0.49\textwidth]{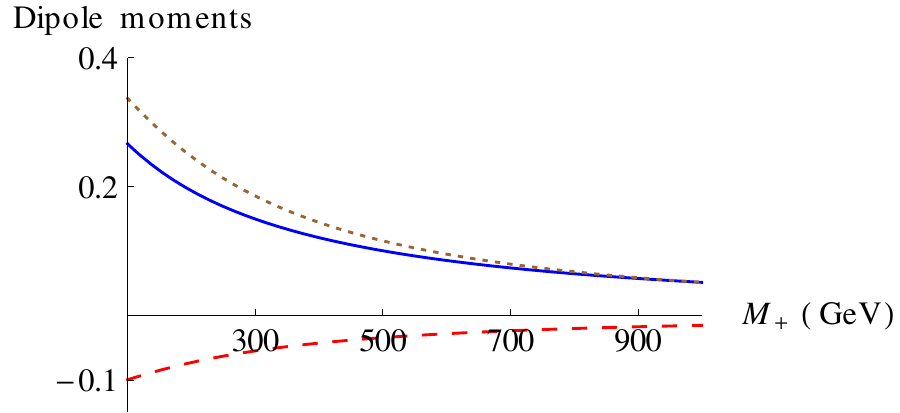}   
 \includegraphics[width=0.49\textwidth]{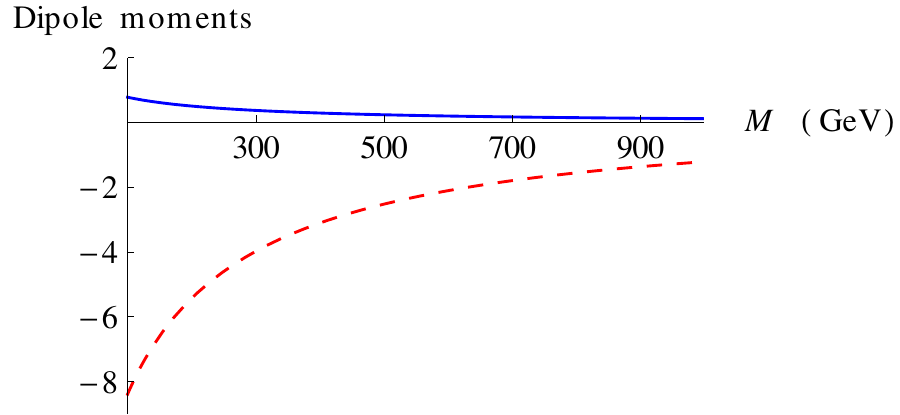}
\caption{
 The dimensionless dipole moments  ${\delta}_d$,
  ${\tilde\delta}_d$, and ${\delta}_W$, defined in Eq.~\eqref{2hdm:hatd}, at the scale $\Lambda_\chi$
  as functions of the charged Higgs-boson mass $M_+$ (left plot) and
  of the mass $M$ of the neutral Higgs bosons $H$ and $A$.
  The blue (solid) and red (dashed) lines correspond to  ${\delta}_d$ and ${\tilde\delta}_d$ (we take $g_s>0$), respectively,
 and  the brown (dotted) line corresponds to  ${\delta}_W$. For the parameter specifications in Eq.~\eqref{assumptions} there
   is no contribution to   ${\delta}_W$ from the neutral Higgs bosons.
}
 \label{Contributions}
\end{figure}

As already mentioned above, the lightest neutral Higgs boson $h$ is to
be identified with the 126 GeV spin-zero boson discovered at the LHC.
The experimental analysis of this resonance does not (yet) prove that
 it is a pure scalar, but the
 data indicate \cite{SANDERS:2013xqa,MEOLA:2013lsa}
that  a possible pseudoscalar component of this state must be smaller than the scalar
one. Therefore, we make the following simplifying assumptions:
\bea
(i)&\qquad  R_{11}\rightarrow 1\,\,, \qquad R_{12}\rightarrow0\,\,,\qquad R_{13}\rightarrow 0\,\,,&\nn\\
(ii)&\qquad M_H\rightarrow M\,\,,\qquad M_A\rightarrow M\,\,.& 
\label{assumptions}
\eea
{Assumption (i) amounts to assuming that the
 lightest Higgs boson $h$ is a pure scalar\footnote{The recent
 papers \cite{Inoue:2014nva,Cheung:2014oaa} analyze EDMs in a 2HDM with a $Z_2$ symmetry and a Higgs potential that
 softly breaks this symmetry and violates $\CP$. They
 take into account the possibility that the $126$ GeV resonance has a
 (small) $\CP$-odd component.}, 
  while with (ii) we assume
 that the two heavier neutral Higgs bosons $H$ and $A$ are (nearly) mass-degenerate.
 These assumptions are not meant to single out a particular
 phenomenologically  or theoretically favored version of the
 a2HDM. They just serve to  simplify the dependence of the quark
 and gluon (C)EDMs on unknown parameters of the model. In this way
 their sizes can be compared. \\

 We can now construct the
  relevant $\CP$-violating  operators up to dimension-six that are generated at a
  high scale $\mu \sim$ a few hundred GeV. Details of our analysis, which essentially follows Ref.~\cite{Jung:2013hka}, are given in Appendix~\ref{A2HDMAppendix}. With the specifications in Eq.~\eqref{assumptions} it turns out that the dominant operators are the EDM and CEDM of the $d$ 
 quark, generated by  two-loop Barr-Zee diagrams \cite{Barr:1990vd} as  shown in Fig.~\ref{diagrams}(a,b),
and the Weinberg operator which is generated by diagrams Fig.~\ref{diagrams}(c) with the
 exchange of a charged Higgs boson. The resulting  (C)EDM of the $d$ quark and the gluon CEDM $d_W$ are
  given in Eqs.~\eqref{2HDMemt}, \eqref{2HDMCmt}, and \eqref{2HDMWmb}, respectively. 
These three dipole moments depend, apart from the unknown
  Higgs boson masses $M$ and $M_+$,  on the common unknown factor  $\text{Im}\, (\varsigma_u^*\varsigma_d)$
 that signifies non-KM $\CP$ violation.
Using the renormalization-group equation for these dipole interactions \cite{Wilczek:1976ry,BraatenPRL, Degrassi:2005zd, Hisano3, Dekens:2013zca} we obtain the following $\slashPT$ effective Lagrangian at the scale $\mu=\Lambda_\chi$ :
\bea  \label{2HDM1GeV}
\vL_{\slashPTsub}   =  -\frac{d_d(\Lambda_\chi)}{2}  \bar d i\simu \ga_5 d\, F_{\mu\nu} -\frac{{\tilde d}_d(\Lambda_\chi)}{2} \bar d i\simu \ga_5 t_a d\, G^a_{\mu\nu}+ \frac{d_W(\Lambda_\chi)}{6} f_{abc}\varepsilon^{\mu\nu\al\bt}G^a_{\al\bt}G^b_{\mu\rho}G_{\nu}^{c \, \rho}\,\,\, .&
\eea
In order to present the sizes of these dipole moments, we define dimensionless quantities ${\delta}_d$,
  ${\tilde\delta}_d$, and ${\delta}_W$ by
 \begin{equation}\label{2hdm:hatd}
d_d(\Lambda_\chi) = e {\delta}_d \frac{{\bar m}\, \text{Im}\, (\varsigma_u^*\varsigma_d)}{v^2} \cdot 10^{-4} \, ,
 \quad 
{\tilde d}_d(\Lambda_\chi)= {\tilde\delta}_d \frac{{\bar m}\, \text{Im}\, (\varsigma_u^*\varsigma_d)}{v^2} \cdot 10^{-4} \, ,
 \quad d_W(\Lambda_\chi)={\delta}_W \frac{\text{Im}\, (\varsigma_u^*\varsigma_d)}{v^2}\cdot 10^{-4} \, ,
\end{equation}
where ${\bar m}(\Lambda_\chi)=4.8$ MeV~\cite{pdg:2012}.
 The dimensionless moments are given as functions of the mass $M_+$ of $H^\pm$ and of the mass $M$ of
  the neutral Higgs bosons $H$ and $A$  in Fig.~\ref{Contributions}.}

 The plots in Fig.~\ref{Contributions}
  show that  the parameter specifications in Eq.~\eqref{assumptions}  imply that the dominant contribution to the $d$-quark EDM and to the gluon CEDM $d_W$ is due to charged Higgs boson exchange.
 In contrast, the contribution of $H^\pm$  to the $d$-quark CEDM is induced through  
 renormalization-group running and is very small as compared to the  contribution of the Higgs bosons $H$ and $A$ to ${\tilde d}_d$.
  The neutral Higgs bosons  also contribute to  the $d$-quark EDM but not to $d_W$, however, the largest part of $d_d(\Lambda_\chi)$ arises when the $d$-quark CEDM is renormalization-group evolved to lower energies.

 For approximately equal masses of $H, A$, and
  the charged Higgs bosons, $M{=}M_H{\simeq}M_A{\simeq}M_{+}$, we find (numerically) the following Higgs-mass independent relations between the dipole moments:
\bea\label{hierarchy}
\tilde d_d(\MQCD) \simeq -7d_d(\MQCD)/e \simeq -20 \bar m d_W(\MQCD)\,\,\,.
\eea
Of course, these relations apply only to a very small region
  in the parameter space of the model.
 Different hierarchies could be realized. 
 For example, if the charged Higgs boson is significantly lighter than the neutral ones, 
 the dominance of the $d$-quark CEDM is reduced. 
  However, since there is no good reason to assume $M_{+} \ll M_H \approx M_A$, we will use Eq.~\eqref{hierarchy} in what follows. 
 Other hierarchies can be studied in similar fashion. Finally, we recall that in non-supersymmetric 2HDMs with the symmetry breaking scale set by the electroweak
  scale $v$, the masses of the Higgs bosons cannot exceed $\sim 1$ TeV.

\subsection{The MSSM}
\label{suse:MSSM}

The minimal supersymmetric extension (MSSM) of the Standard Model is another popular SM extension that
 is theoretically well motivated and  allows for non-KM $\CP$ violation.  In general, the soft supersymmetry (SUSY) breaking terms
  contain many $\CP$-violating phases. Often the analysis of $\CP$-violating effects is restricted
  to versions of the MSSM that contain, apart from the KM phase, two additional reparametrization invariant phases, which
 are usually chosen to be the phase of the $\mu$ term
   and a common phase of the trilinear fermion-sfermion-gaugino
  interactions. Other SUSY $\CP$ scenarios  are also discussed. For a review, see Ref.~\cite{Ibrahim:2007fb}.
  
It has been known for a long time \cite{Ellis:1982tk,Buchmuller:1982ye,Polchinski:1983zd,delAguila:1983kc}
 that SUSY $\CP$ phases generate EDMs of quarks and charged leptons and CEDMs of quarks
  at one loop by the exchange of charginos, neutralinos, and gluinos between sfermions and fermions. The Weinberg operator
  is induced at two loops by diagrams involving gluinos, squarks, and quarks \cite{Dai:1990xh}.
  Moreover, loop-induced SUSY threshold corrections \cite{Lebedev:2002ne}  and, in addition,
 sizable mixing between the two $\CP$-even and the $\CP$-odd
  neutral Higgs bosons at one loop \cite{Pilaftsis:1998dd}, which can occur for a certain set of values in the SUSY parameter space,
  lead to $\CP$-violating effects due to neutral
  Higgs boson exchange. The latter effect induces two-loop (nominally three-loop) Barr-Zee type contributions to the EDMs of quarks and leptons and to the CEDMs
  of quarks \cite{Chang:1998uc,Chang:1999zw,Pilaftsis:2002fe,Carena:2000ks} and  additional two-loop contributions to the Weinberg operator.   Two-loop rainbow-like contributions to the (C)EDM of quarks were analyzed in \cite{Pilaftsis:1999bq,Yamanaka:2012ia}.
  There is a huge literature on SUSY-induced (C)EDMs  that includes Refs.~\cite{Kizukuri:1992nj,Ibrahim:1998je,Brhlik:1998zn,Bartl:1999bc,Abel:2000hn,Abel:2001vy,Lebedev:2002ne,Demir:2003js,ArkaniHamed:2004yi,Abel:2005er,Degrassi:2005zd,Chang:2005ac,Giudice:2005rz,Hisano:2006mj,YaserAyazi:2006zw,Ellis:2008zy,Li:2010ax,Ellis:2011hp,Fuyuto:2013gla}.

Because the SUSY-induced (C)EDMs of quarks and leptons are quite
large,  the SUSY particles must be heavy and/or the SUSY phases have
to be small in order that the model does not get in conflict with
 the existing experimental bounds on various EDMs (see below), or cancellations between the various contributions to an atomic/electron EDM and to the
 EDM of the neutron must occur
 \cite{Ibrahim:1998je,Bartl:1999bc,Brhlik:1998zn,YaserAyazi:2006zw}. 
 Another possibility is that flavour-blind SUSY phases are
 absent and $\CP$ violation is associated with the SUSY Yukawa
 interactions, which leads to small EDMs not in conflict with
 experimental bounds (\textit{cf}., e.g., Ref.~\cite{Abel:2000hn}).
 
 Non-observation of SUSY signatures
  at the LHC  so far leads to the conclusion that
  most of the SUSY particles must be quite heavy if they exist. The interpretation of the LHC data depends of course on the specific SUSY scenario that is used for the data analysis.
 For instance, the recent global fit \cite{Bechtle:2012zk,Bechtle:2013mda} indicates that the masses of the first and second generation squarks and of the gluinos are above
  $\sim 2$ TeV, while the masses of the charginos, neutralinos, and third generation squarks are $\gtrsim 500$ GeV. The lightest of the neutral Higgs bosons is to be identified with
  the 126 GeV resonance, while the two heavier states $H$ and $A$ and
  the charged Higgs boson are above 1 TeV. 
 In scenarios that are in accord with these fit results
 and  assume that the SUSY breaking scale is significantly higher than
 the electroweak scale, the SUSY phases are 
 nevertheless constrained by the experimental EDM bounds \cite{Fuyuto:2013gla}.

A comprehensive compilation and analysis of SUSY-induced EDMs of the
neutron, deuteron, the Thallium (Tl) and Mercury (Hg) atoms was made
in Ref.~\cite{Ellis:2008zy}.
 (A similar analysis, taking into account more independent SUSY $\CP$
 phases, was performed in Ref.~\cite{Li:2010ax}.)
 From this analysis follows that several SUSY $\CP$ scenarios
 \cite{Carena:2000ks,Ellis:2004fs,Ellis:2007kb} induce, for
 valence $u$ and $d$ quarks,  
  the following $\slashPT$ low-energy effective quark-gluon and photon
  Lagrangian at the scale
 $\mu=\Lambda_\chi$:
\begin{equation}  \label{SUSY1GeV}
\vL_{\slashPTsub}   =  - \frac{1}{2}\sum\limits_{q=u,d} \left( d_q(\Lambda_\chi)  \bar q i\simu \ga_5 q\, F_{\mu\nu} 
 + {\tilde d}_q(\Lambda_\chi) \bar q i\simu \ga_5 t_a q\, G^a_{\mu\nu} \right)
 + \frac{d_W(\Lambda_\chi)}{6} f_{abc}\varepsilon^{\mu\nu\al\bt}G^a_{\al\bt}G^b_{\mu\rho}G_{\nu}^{c \, \rho}\, .
\end{equation} 
In addition, $\CP$-violating four-quark operators generated by loop-induced Higgs-boson mediated $\CP$ violation
  can be generated.    Such operators can also be induced by one-loop box diagrams involving SUSY particles  \cite{Demir:2003js}, but these are  subdominant effects.

The magnitudes and signs of the (C)EDMs of the $u$ and $d$ quarks and of the gluon CEDM in Eq.~\eqref{SUSY1GeV}  depend on the masses of the
 SUSY particles, the sfermion-fermion-gaugino mixing matrix elements,
 and on the SUSY $\CP$ phases. For certain sets of SUSY scenarios that
 are phenomenologically viable, the low-energy effective Lagrangian can be further specified. These scenarios include
\begin{itemize}
\item[a)] Heavy SUSY spectrum with a common mass scale $ > 1$ TeV and
  rather large $\tan\beta= v_2/v_1$. The global fit \cite{Bechtle:2013mda}
  is in accord with $5 \lesssim \tan\beta \lesssim 30$. In this case
    the one-loop contributions to the $d$-quark (C)EDM dominate in
    Eq.~\eqref{SUSY1GeV}, \textit{cf}., for instance, Ref.~\cite{Abel:2005er}.
  As to  $\CP$-violating four-quark operators: In view of $\tan\beta \lesssim 30$,
  of the negative experimental results on SUSY signatures,
  and of the presently known properties of the 126 GeV resonance (see Sect.~\ref{2HDM}) we conclude that the  coefficients of such operators are rather small.

\item[b)] Heavy first and second generation sfermions, $m_{\tilde f}>
  10$ TeV. In this case the  contributions to the quark and
   electron (C)EDMs are suppressed \cite{Kizukuri:1992nj}
  and the dominant contribution to Eq.~\eqref{SUSY1GeV} is due to the
  Weinberg operator \cite{Abel:2005er}.
\item[c)] `Split SUSY' \cite{ArkaniHamed:2004yi}. Here also the 
 third sfermion generation is very heavy, so that sfermions and gluinos decouple
  from physics at the electroweak scale. In this case the one- and
  two-loop quark CEDMs and the coefficient of the Weinberg operator are
  tiny; i.e., the   low-energy effective Lagrangian
  in Eq.~\eqref{SUSY1GeV} contains only the EDMs of the $u$ and $d$ quark that are
  generated, like the EDM of the electron, by two-loop Barr-Zee type
  diagrams that involve charginos and neutralinos
  \cite{ArkaniHamed:2004yi,Giudice:2005rz}. In this scenario there is 
  a strong correlation between the magnitudes of the electron and 
   neutron  EDM  \cite{Giudice:2005rz}.
\end{itemize}

In summary, several scenarios are possible within the MSSM.
  However, it appears most natural to us to consider  in our analysis of
the following sections all $\slashPT$ operators of Eq.~\eqref{SUSY1GeV} -- which is
then similar to the a2HDM. Therefore, in the discussions
below we discuss the a2HDM in more
detail and only briefly remark at the end of the corresponding sections on the MSSM.

\section{The chiral Lagrangian} 
\label{chiral}

Below the energy scale $\Lambda_{\chi}\sim 1$ GeV nonperturbative techniques are required to describe hadronic interactions.
The degrees of freedom of the effective field theory of QCD for the two\footnote{In this work we do not consider the strange quark explicitly which can be done by using $SU(3)$ chiral
perturbation theory. We do not expect that effects of dynamical strange quarks exceed the uncertainties given
below.} valence quark flavors, 
$SU(2)$ \textit{Chiral Perturbation Theory} ($\chi$PT), are pions ($\pi$), nucleons ($N$) and photons ($A_{\mu}$) 
(see, e.g., Ref.~\cite{Gasser:1983yg,BKM95,Gasser:1987rb,Weinberg:1978kz,Weinbergbook}). The pions are the Goldstone bosons associated with the spontaneous chiral 
symmetry breakdown $SU(2)_L\!\times\! SU(2)_R\rightarrow SU(2)_V$. The chiral $SU(2)_L\!\times\!SU(2)_R$ symmetry is only approximate 
and is explicitly broken by the finite quark masses, the quark charges and, in our case, the effective $\slashPT$ operators.

The $\chi$PT Lagrangian contains all interactions between these fields which are allowed by the symmetries of QCD. 
Chiral-invariant interactions involving pions always appear with a derivative acting on the pion field and it is 
this property which gives $\chi$PT its consistent power counting. Interactions involving the pion field without derivatives 
are induced by the chiral-symmetry-breaking interactions in the QCD Lagrangian and are proportional 
to the small chiral-breaking parameters. This explains the relative lightness of the pion whose mass is proportional to the small average quark mass $\mpi^2 \sim \bar m $. 

We are interested in the effects of the $\slashPT$ operators appearing in the various scenarios discussed in Sect.~\ref{3scenarios}. 
 They all induce $\slashPT$ hadronic interactions, but the form of these interactions differs for each scenario because the various $\slashPT$ operators transform differently under chiral 
and isospin rotations. Each interaction is accompanied by a low-energy constant (LEC) determined by nonperturbative physics. In most cases considered here, these LECs are unknown 
and, barring lattice QCD calculations, need to be estimated by a model calculation. Reasonable estimates can be obtained by QCD sum rules \cite{Pospelov_review} or 
naive dimensional analysis (NDA)~\cite{NDA,Weinberg:1989dx}. 
In the case of the $\tb$ term the LECs can be controlled quantitatively, as we will discuss.

The hadronic Lagrangians for the dimension-four and -six operators have been constructed in Refs.~\cite{BiraEmanuele, deVries:2012ab}. 
In the next sections 
 we discuss which hadronic operators 
 are induced by the low-energy effective Lagrangians  Eqs.~\eqref{QCD2}, \eqref{mLRSM1GeV}, \eqref{2HDM1GeV}, and  \eqref{SUSY1GeV}. 
  Among the most important of these hadronic operators are the $\slashPT$ pion-nucleon interactions, 
which provide long-range $\slashPT$ forces between nucleons. We discuss the $\slashPT$ pion-nucleon 
vertices first. Afterwards we will study the EDMs of the neutron and proton which are not 
only interesting by themselves but also are an important ingredient 
  of light-nuclear EDMs, to be discussed in subsequent sections.  

\subsection{Parity- and time-reversal-odd pion-nucleon interactions} 
Interactions between pions and nucleons that break $P$ and $T$ play an important role
 in the calculation of the EDMs of nucleons and nuclei because they induce long-range $\slashPT$ forces. Historically, hadronic EDMs have often been discussed in terms of a one-boson-exchange model in which it is assumed that $P$ and $T$ violation is induced by the following two\footnote{In phenomenological studies a third interaction $\bar g_2 \Nb \pi_3 \tau_3 N$ is often included as well, but this interaction only appears at 
higher orders than those considered here for all dimension-four and -six $\slashPT$ 
operators \cite{BiraEmanuele, deVries:2012ab}.} nonderivative interactions:
\begin{equation}\label{g01}
\mathcal L = \bar g_0 \Nb \boldpi\cdot \boldtau N+ \bar g_1 \Nb \pi_3 N \,\,\,.
\end{equation} 
We discuss below for each scenario the (relative) sizes of the $\bar g_i$ and whether additional interactions should be taken into account.  

\subsubsection{The $\tb$ scenario}
The $\tb$-term is the isospin-conserving  element of the same  
chiral-symmetry breaking quark-antiquark multiplet as the isospin-breaking component of the quark mass matrix --
 both terms are connected by an axial $SU(2)$ rotation. Therefore, all terms in the effective Lagrangian induced by the $\tb$-term are linked to terms arising from the so-called {\em strong } isospin breaking, 
 i.e., isospin-breaking resulting from the strong interactions \cite{CDVW79, BiraEmanuele}. The induced leading-order term in the pion-nucleon sector of the effective Lagrangian is proportional 
to the quark-mass induced part of the proton-neutron mass difference $(m_n -m _p)^{\mathrm{strong}}$  \cite{FMS98}. It gives the leading contribution to the coupling constant $\bar g_0^{\theta}$ \cite{Pospelov_deuteron,BiraEmanuele, Jan_2013},
\begin{equation}\label{g0theta}
\bar g_0^\theta = \frac{(m_n -m _p)^{\mathrm{strong}} (1-\varepsilon^2)}{4\Fp \varepsilon} \tb = (-0.018 \pm 0.007)\,\tb\,\,\,,
\end{equation}
where $(m_n -m _p)^{\mathrm{strong}} = (2.6 \pm0.85)$ MeV  \cite{Gasser:1982ap,Walker-Loud}, $\varepsilon \equiv(m_u{-}m_d)/(m_u{+}m_d)= -0.35\pm0.10$ \cite{pdg:2012}, and
$\Fp = 92.2$ MeV have been used.

The contributions to the coupling constant $\bar{g}_1^{\theta}$ induced by the $\tb$-term 
can be traced back 
to the emergence of pion-tadpole terms in the pion-sector Lagrangian, which have to be removed by field 
redefinitions \cite{BiraEmanuele,Jan_2013}. These field redefinitions generate 
the leading contribution to $\bar{g}_1^{\theta}$ given by \cite{Pospelov_deuteron,BiraEmanuele,Jan_2013}:
\begin{equation}\label{g1theta}
\bar g_1^\theta = \frac{2 c_1 (\delta \mpi^2)^{\mathrm{strong}}(1-\varepsilon^2)}{\Fp \varepsilon} \tb = (0.003\pm0.002)\tb\,\,\,,
\end{equation}
where the LEC $c_1 =( -1.0\pm0.3)\,\mathrm{GeV}^{-1}$~\cite{pid} is related to the nucleon $\sigma$-term and where $(\delta \mpi^2)^{\mathrm{strong}}\simeq (\varepsilon \mpi^2)^2/(4(m_K^2-m_{\pi}^2))$ \cite{Gasser:1984gg}
 is the strong contribution to the square of the mass splitting between charged and neutral pions. 
The uncertainty in Eq.\,(\ref{g1theta}) has been increased to account for the contribution to $\bar g_1^\theta$  
 by another independent term \cite{BiraEmanuele} in the next-to-next-to-leading-order pion-nucleon Lagrangian \cite{Fettes:2000gb} 
with an  LEC of unknown strength,
 which here has  been conservatively bounded by its NDA estimate. 
In fact, 
a calculation based on resonance saturation  predicts  only one third of this estimate as an upper bound~\cite{Jan_2013}.

In summary,
the coupling constant  $\bar{g}_1^{\theta}$ is suppressed with respect to the coupling constant $\bar{g}_0^{\theta}$ by the ratio $\bar g_1^\theta/\bar g_0^\theta = -0.2\pm0.1$~\cite{Jan_2013}. 
This suppression, however, is less than the NDA prediction $|\bar g_1^\theta/\bar g_0^\theta|=\mathcal O(\varepsilon \mpi^2/\Lambda_\chi^2) \simeq 0.01$~\cite{BiraEmanuele}.

\subsubsection{The minimal left-right symmetric scenario}\label{piNmLRSM}

As discussed in Sect.~\ref{mLRSM}, in the mLRSM the most important $\slashPT$ contributions at low energies are due to the four-quark interactions in Eq.~\eqref{mLRSM1GeV}. 
The chiral Lagrangian induced by these operators has been constructed in Ref.~\cite{deVries:2012ab} 
and we recall
 the main results here. First of all, 
 the two four-quark interactions have the same chiral properties and induce hadronic interactions of identical form (although the LECs, of course, will be different).  We therefore use $\mathrm{Im}\,\Xi$ to collectively denote $\eta_{1,8}\, \mathrm{Im}\big(V^{ud\,*}_L\, \Xi_{1}\big)$ and denote the associated four-quark operator as the four-quark left-right operator (FQLR).

The FQLR is a chiral- and isospin-breaking interaction,
 however it does not transform under chiral symmetry as any term in the conventional QCD Lagrangian, but instead transforms in a more complicated fashion \cite{deVries:2012ab}. 
The $\slashPT$ LECs induced by the FQLR are therefore not linked to any strong LECs as was the case for the $\tb$ term. 
 Unfortunately, this implies that we need to resort to different techniques to estimate the LECs. 
 We will use NDA because
other methods, such as QCD sum rule estimates are, to our knowledge, 
not available. Nevertheless, even without 
 detailed knowledge of the LECs, considerations based on chiral symmetry give a lot of information 
on the hierarchy of the hadronic interactions. 

Because the FQLR violates isospin symmetry it does not contribute to $\bar g_0^{\mathrm{LR}}$ directly. Instead, it contributes to isospin-violating LECs in two ways. 
First of all, a pion tadpole is induced. However, due to the complicated chiral properties of the FQLR, this tadpole is associated with a three-pion 
 vertex  \cite{deVries:2012ab} which, in the so-called $\sigma$-parametrization of $SU(2)$ $\chi$PT,  see  e.g., Ref.~\cite{Hanhart_Wirzba}, reads
\begin{equation}
\mathcal L^{\mathrm{LR}} = \bar \Delta^{\mathrm{LR}}\Fp \pi_3\left(1-\frac{\boldpi^2}{2\Fp^2}\right)\,\,\,.
\end{equation}
In addition, the FQLR induces a direct contribution to $\bar g_1^{LR}$. The sizes of the LECs can be estimated by NDA: 
\begin{equation}\label{NDA_g1_mLRSM}
|\bar \Delta^{\mathrm{LR}}| = \Or(\Fp^2 \Lc^2\mathrm{Im}\, \Xi) \simeq (0.01\,\mathrm{GeV}^4)\mathrm{Im}\, \Xi\,\,,\,\,\, |\bar g_1^{\mathrm{LR}}|=\Or(\Fp \Lc \mathrm{Im}\, \Xi)\simeq (0.1\,\mathrm{GeV}^2) \mathrm{Im}\, \Xi\,.
\end{equation}
Just as for the $\tb$ term, the tadpole can be removed using field redefinitions.
 However, differently from the $\tb$ term,
 a three-pion vertex is left behind (see Eq.~\eqref{chiralmLRSM}). Moreover, the removal of the tadpole induces an 
 additional contribution to $\bar g^{\mathrm{LR}}_1$  proportional to $\bar \Delta^{\mathrm{LR}}$:
\begin{equation}\label{g1total}
\bar g^{\mathrm{LR}}_1 \rightarrow \bar g^{\mathrm{LR}}_1+ \bar g_1^{\mathrm{LR}\,\prime}\,\,,\qquad \bar g^{\mathrm{LR}\,\prime}_1 = -\frac{4 c_1 \bar \Delta^{\mathrm{LR}}  }{\Fp}\,\,\,.
\end{equation}
Since $c_1 =\Or(1/\Lc)$, the additional contribution is formally of the same order as the original term.
 However, numerically it might be somewhat larger because $4c_1$ is bigger than expected from NDA. 
In addition, a first non-vanishing contribution to $\bar g_0^{\mathrm{LR}\,\prime}$ appears\footnote{If
 the $\tb$ term is removed by the Peccei-Quinn mechanism, this contribution to $\bar g_0^{\mathrm{LR}\,\prime}$ would be absent \cite{Emanueleprivate}. Because
  it is small anyway this has no consequences for our analysis. }, also proportional to $\bar \Delta^{\mathrm{LR}}$:
\begin{equation}
\bar g^{\mathrm{LR}\,\prime}_0 =  - \frac{(m_n -m _p)^{\mathrm{strong}}\bar \Delta^{\mathrm{LR}} }{2\mpi^2 \Fp}\,\,\,.
\end{equation}

In conclusion, the relevant pionic and pion-nucleon interactions are given by
\begin{equation}\label{chiralmLRSM}
\mathcal L^{\mathrm{LR}} = -\bar \Delta^{\mathrm{LR}} \frac{\pi_3 \boldpi^2}{2\Fp} + \bar g^{\mathrm{LR}\,\prime}_0 \Nb \boldpi\cdot \boldtau N+ (\bar g^{\mathrm{LR}}_1 + \bar g^{\mathrm{LR}\,\prime}_1) \Nb \pi_3 N\,\,\,.
\end{equation}
Because $\bar g^{\mathrm{LR}\,\prime}_0$ and $\bar g^{\mathrm{LR}\,\prime}_1$ both depend on the same LEC $ \bar \Delta^{\mathrm{LR}}$, their ratio depends only on known quantities: 
\begin{equation}\label{ratiomLRSM}
\frac{\bar g^{\mathrm{LR}\,\prime}_0}{\bar g^{\mathrm{LR}\,\prime}_1} =\frac{(m_n -m _p)^{\mathrm{strong}} }{8 c_1 \mpi^2} = -0.02\pm0.01\,\,\,.
\end{equation}
Unless $\bar g^{\mathrm{LR}}_1$ and $\bar g^{\mathrm{LR}\,\prime}_1$ cancel to a high 
degree --  which is not expected on any grounds -- 
 the coefficient $\bar g^{\mathrm{LR}\,\prime}_0$ is much smaller \cite{deVries:2012ab} than the combination $\bar g^{\mathrm{LR}}_1 + \bar g^{\mathrm{LR}\,\prime}_1$ which appears in observables. From now on, we relabel $\bar g^{\mathrm{LR}}_1 + \bar g^{\mathrm{LR}\,\prime}_1 \rightarrow \bar g^{\mathrm{LR}}_1$ and $\bar g_0^{\mathrm{LR}\,\prime} \rightarrow \bar g_0^{\mathrm{LR}}$, and take $|\bar g^{\mathrm{LR}}_0/\bar g^{\mathrm{LR}}_1| \ll 1$. This result is in stark contrast with the $\tb$ scenario where $\bar g_0^{\tb}$ is the dominant interaction. This difference has important consequences for light-nuclear EDMs.

\subsubsection{The a2HDM and MSSM scenarios}\label{piN2HDM}
Next we discuss the hierarchy of pion-nucleon interactions which emerges in the scenario of Sects.~\ref{2HDM} and \ref{suse:MSSM}. In the a2HDM and MSSM scenarios we 
must compare the contributions from the three operators appearing in Eqs.~\eqref{2HDM1GeV} and \eqref{SUSY1GeV}. We focus first on the a2HDM scenario because here we have some idea on the relative sizes of the quark (C)EDMs and the Weinberg operator. After discussing the a2HDM, we consider briefly how our findings might be altered in the MSSM.

The contributions from the qEDM to the pion-nucleon LECs are highly suppressed because of the appearance of the photon which needs to be 
integrated out. As such, the qEDM contributions are suppressed by the typical factor $\alpha_{\mathrm{em}}/\pi \sim 10^{-3}$ and can be safely neglected. Because  the Weinberg operator  conserves chiral symmetry, it
 cannot  directly induce the
 \mbox{$\slashPT$} pion-nucleon couplings which break chiral symmetry \cite{Pospelov_Weinberg}. Instead, 
an additional insertion of the quark mass (difference) is required which implies, by NDA, that the LECs scale as $|\bar g^{\mathrm{H}}_0| = \Or (\bar m \Lc\, d_W)$ 
and $|\bar g^{\mathrm{H}}_1| = \Or (\varepsilon \bar m \Lc\, d_W)|$~\cite{deVries2010a}.  
On the other hand, the down quark CEDM in Eq.~\eqref{2HDM1GeV} can induce the pion-nucleon couplings 
directly such that, for this source,  $|\bar g^{\mathrm{H}}_{0,1}| = \Or ( \Lc\, \tilde d_d)$ \cite{deVries2010a}. 
Together with the observation that $\bar m |d_W|$ is about an order of magnitude smaller than $|\tilde d_d|$ in the model under investigation, we conclude that the pion-nucleon couplings are dominated by the qCEDM.

To check whether the NDA estimate is reasonable we compare it to results obtained in Refs.~\cite{Pospelov_piN, Pospelov_review} 
where the pion-nucleon LECs were investigated in the framework of QCD sum rules. 
It was found that\footnote{It should be noted that these results assume a 
Peccei-Quinn mechanism to remove the $\tb$ term. This also shifts the qCEDM contributions to pion-nucleon LECs. 
However, the order of magnitude stays the same. Since we use these results as a sanity check of the NDA estimate, this poses no real problem. }
\begin{equation}\label{g01a2hDM}
 \bar g^{\mathrm{H}}_1 = -(2^{+4}_{-1}\,\mathrm{GeV} )\, \tilde d_d\,\,,\qquad 
\bar g^{\mathrm{H}}_0 \simeq (0.5\pm 1)\,\mathrm{GeV}\, \tilde d_d\,\,\,.
\end{equation}
The coupling
 $\bar g^{\mathrm{H}}_1$ is somewhat bigger than the NDA estimate but not in disagreement. 
The calculation of $\bar g^{\mathrm{H}}_0$ has a relatively larger uncertainty (even an uncertain sign) which 
is also harder to quantify \cite{Pospelov_piN}. The size of $\bar g^{\mathrm{H}}_0$ is somewhat smaller than $\bar g^{\mathrm{H}}_1$,  and it is in agreement with NDA. 

Considering the large uncertainties in these estimates, from now on we will take for the a2HDM scenario that $|\bar g^{\mathrm{H}}_0| \simeq |\bar g^{\mathrm{H}}_1|$ as indicated by NDA. However, there is a significant uncertainty involved and the only way to improve the situation is, most likely, a direct lattice calculation. 

The situation in the MSSM is similar. Unless the Weinberg operator is larger than the $u$ and $d$ quark CEDMs, also in this scenario the pion-nucleon LECs are dominated by the qCEDM contributions. The $u$-quark CEDM is now expected to be significant as well which means that the result in Eq.~\eqref{g01a2hDM} should be slightly altered: $\tilde d_d$ in the expression for $\bar g_1$ ($\bar g_0$) should be replaced by $\tilde d_d - \tilde d_u$ ($\tilde d_d + \tilde d_u$). Nevertheless, we expect again that the pion-nucleon LECs are of similar size, with the possibility that $|\bar g^{\mathrm{MSSM}}_1|$ is slightly larger than $|\bar g^{\mathrm{MSSM}}_0|$. 
Even in the case that the Weinberg operator is much larger than the quark CEDMs, a large hierarchy between $\bar g^{\mathrm{MSSM}}_0$ and $\bar g^{\mathrm{MSSM}}_1$ is not expected to appear. The NDA estimates given above tell us that the LECs are of similar size, apart from a possible small suppression of $\bar g^{\mathrm{MSSM}}_1$ due to insertion of the quark-mass difference $\varepsilon$. In what follows we therefore take $|\bar g^{\mathrm{MSSM}}_0| \simeq |\bar g^{\mathrm{MSSM}}_1|$.

\subsection{The EDMs of the neutron and proton}
Now that we have discussed the $\slashPT$ pion-nucleon interactions for each of the scenarios we turn to the calculation of the nucleon EDMs. The nucleon EDMs
 obtain contributions
 from one-loop diagrams involving the $\slashPT$ pion-nucleon
 vertices in Eq.~\eqref{g01}. 
 However, these diagrams are ultraviolet-divergent
  and renormalization requires counterterms to absorb these divergences and the associated scale dependence. Such counterterms appear 
naturally in $\chi$PT in the form of $\slashPT$ nucleon-photon vertices \cite{Pich:1991fq, Borasoy:2000pq,BiraHockings, Narison:2008jp,ottnad,deVries2010a, Mer11,  Guo12},
\begin{equation}\label{Ngamma}
\mathcal{L}_{N\gamma}  = 
 - 2\, \Nb\left(\bar{d}_{0}+\bar{d}_{1}\tau_{3}\right)S^{\mu}N\, v^{\nu}\Fmu\,\,\,,
\end{equation}
in terms of the nucleon spin $S^\mu = (0\,,\,\vec \sigma/2)^T$ and velocity $v^\mu = (1\,,\, \vec 0)^T$ in the nucleon restframe. These counterterms 
appear in all scenarios discussed here, but their sizes, of course, vary depending on the scenario under investigation. 

Before discussing the new LECs $\bar d_{0,1}$, let us first discuss the calculation of the nucleon EDM in 
terms of $\slashPT$ interactions in Eqs.~\eqref{g01} and \eqref{Ngamma}\footnote{Contributions from the three-pion vertex $\bar \Delta^{\mathrm{LR}}$ 
in Eq.~\eqref{chiralmLRSM} to the nucleon EDM vanish up to next-to-leading order \cite{deVries:2012ab}.}. This calculation has been performed in 
\mbox{$SU(2)_L\times SU(2)_R$} heavy-baryon $\chi$PT up to next-to-leading order in Refs.~\cite{BiraHockings, deVries2010a, Mer11} (for 
\mbox{$SU(3)_L\times SU(3)_R$} results, see Refs.~\cite{Borasoy:2000pq,ottnad, Guo12}) and gives
 for the neutron ($d_n$) and proton EDM ($d_p$)
\bea \label{nucleonEDM}
  d_n& = & {\bar d}_0 -\bar d_1-\frac{e g_A\bar g_0}{8\pi^2 F_\pi} \left(  \ln
\frac{m_\pi^2}{m_N^2} -\frac{\pi m_\pi}{2 m_N} \right)\,\,\,,\nn\\
\label{eq:dpfull}
d_p & = & {\bar d}_0 + \bar d_1+\frac{e g_A}{8\pi^2 F_\pi} \left[ \bar g_0 \left(  \ln
\frac{m_\pi^2}{m_N^2} -\frac{2\pi m_\pi}{m_N} \right) -
\bar g_1 \frac{\pi m_\pi}{2 m_N} \right]\,\,\,,
\eea
where $e>0$. Furthermore,
$g_A \simeq 1.27$
is the strong pion-nucleon coupling constant~\cite{pdg:2012}, $m_N$ the nucleon mass, and the divergence has been absorbed into the counterterms. 
The leading loop result reproduces the famous result obtained in Ref.~\cite{CDVW79}, where current algebra techniques were applied.  

The dependence of the nucleon EDMs on the LECs $\bar d_{0,1}$ implies that considerations based on chiral symmetry alone cannot tell us the sizes of these EDMs. Even in the $\tb$ scenario, in which we have relatively precise knowledge of the LECs $\bar g_{0,1}$ (see Eqs.~\eqref{g0theta} and \eqref{g1theta}), the exact dependence of $d_n$ and $d_p$ on $\tb$ is unclear due to the unknown finite parts of $\bar d_{0,1}$.  The same holds, of course, for the other scenarios, where not even the LECs $\bar g_{0,1}$ are known precisely.  Thus, the EDM results in the nucleon case alone are of limited use to get information on the physics that generated them.
The strength
of our methods will become much more visible when two-nucleon contributions, i.e., the cases
of the deuteron and tri-nucleon EDMs, will be investigated in Sect.~\ref{lightnuclei}. 
To get quantitative information on the nucleon EDMs other techniques are necessary.
  Let us now, for each of the scenarios, discuss the sizes of the nucleon EDMs.

\subsubsection{The $\tb$ scenario}
By inserting the values of $\bar g_{0,1}$ from Eqs.~\eqref{g0theta} and \eqref{g1theta} into Eq. \eqref{nucleonEDM} it is possible to evaluate the loop contributions to the nucleon EDMs:
\begin{equation}
d_n^{\tb,\mathrm{loop}} = (-2.5\pm0.9)\cdot 10^{-16}\,\tb\,\ecm\,\,,\qquad d_p^{\tb,\mathrm{loop}} = (2.8\pm0.9)\cdot 10^{-16}\,\tb\,\ecm \,\,.
\end{equation}
This sets the scale for the nucleon EDM but the actual numbers can change due to the LECs $\bar d_{0,1}$. The sizes of the LECs can be estimated by NDA. 
This yields
\begin{equation}
|\bar d_{0,1}^{\,\tb}| = \mathcal O\left(e \tb\, \frac{\mpi^2}{\Lc^3}\right) \simeq 3\cdot 10^{-16}\,\tb\,\ecm\,\,\,, 
\end{equation} 
which is
of similar size as the loop contributions.
By combining a $\chi$PT calculation with lattice QCD data, it recently became possible to compute the total nucleon EDM (loop and tree-level contributions) \cite{Guo12}. 
It was found that\footnote{Lattice data at
lower pion masses, where the chiral extrapolations will be more reliable, are expected to become available soon. A new look at the single-nucleon EDM predictions is work in progress.}
\begin{equation}\label{latticetheta}
d_n^{\tb} = (-2.9\pm0.9)\cdot 10^{-16}\,\tb\,\ecm\,\,,\qquad d_p^{\tb} = (1.1\pm1.1)\cdot 10^{-16}\,\tb\,\ecm\,\,,
\end{equation}
which is the result we will use in what follows. For the deuteron EDM, an important quantity is the sum of the nucleon EDMs which 
is
 \begin{equation}\label{isoscalartheta}
d_n^{\tb} + d_p^{\tb} = (-1.8\pm 1.4)\cdot 10^{-16}\,\tb\,\ecm\,\,,
\end{equation}
with a significant uncertainty. 

\subsubsection{The minimal left-right symmetric scenario}\label{nEDMmLRSM}
In the 
 case of the mLRSM the situation is far more uncertain than 
for the $\tb$ term. 
Because  $\bar g_0^{\mathrm{LR}}$ is significantly suppressed (see Eq.~\eqref{ratiomLRSM}),
 this implies that the loop contributions proportional to $\bar g_0^{\mathrm{LR}}$ are actually subleading. 
Up to next-to-leading order the neutron EDM does not depend on $\bar g^{\mathrm{LR}}_1$, and therefore in Ref.~\cite{Seng14} 
the calculation was extended to next-to-next-to-leading order. It was concluded that both the neutron and proton EDM obtain dominant contributions from the LECs $\bar d_{0,1}$, while the loop contributions are an order of magnitude smaller. Unfortunately, the exact sizes of the EDMs as function of the fundamental mLRSM parameter $\mathrm{Im}\,\Xi$
 is rather uncertain. There exists, to our knowledge, no reliable model calculation.
Therefore
 we cannot do better than NDA, which gives
\begin{equation}\label{NDA_dn_mLRSM}
   \left| d_{n,p} \right| = \Or \left(e\mathrm{Im}\,\Xi \frac{\Fp^2}{\Lc}\right) \simeq (10\,\mathrm{MeV})\, e\mathrm{Im}\,\Xi\,\,\, ,
\end{equation}
with an unknown sign and a large uncertainty in its magnitude.

\subsubsection{The a2HDM and MSSM scenarios}
\label{nEDM2HDM}
Both within the a2HDM and the MSSM, the nucleon EDMs obtain contributions 
 from each of the three operators in Eqs.~\eqref{2HDM1GeV} and \eqref{SUSY1GeV}, respectively. 
 For the pion-nucleon interactions we were fortunate that one of the operators gave
  dominant contributions which simplified the analysis. For the nucleon EDM we do not have this advantage because the qEDM and Weinberg operator 
 induce the tree-level LECs $\bar d_{0,1}$ without additional
suppressions \cite{deVries2010a}. As a consequence, we need to study all three operators. 

To be specific we start the discussion with the a2HDM.
The discussion in Sect.~\ref{piN2HDM} tells us that the loop contributions to the nucleon EDMs are dominated by the qCEDM because the $\slashPT$ pion-nucleon LECs are suppressed for the qEDM and Weinberg operator. Using the NDA estimate for $\bar g_{0}$ (in good agreement with the QCD sum rules result \cite{Pospelov_piN}) gives
\begin{equation}
|d_{n}^{\mathrm{H},\,\mathrm{loop}}| \simeq - |d_{p}^{\mathrm{H},\,\mathrm{loop}}|\simeq  (0.7)\,e\tilde d_d\,\,\,,
\end{equation}
in terms of the down-quark CEDM $\tilde d_d$. The tree-level terms $\bar d_{0,1}$ obtain contributions from all operators. NDA tells us \cite{deVries2010a}:
\begin{equation}
 \left| d_{n,p}^{\mathrm{H},\,\mathrm{tree}} \right|= \Or\left( d_d,\,\,\, \frac{e \tilde d_d}{4\pi},\,\,\, \Fp\,ed_W  \right)\simeq  \left( d_d,\,\,(0.1) \,e\tilde d_d,\,\, (0.1\,\mathrm{GeV})\,e\, d_W\right)\,\,\,,
\end{equation}
which implies  that the loop contributions of the qCEDM  are larger than its tree-level contributions.
The approximate hierarchy of the dipole moments $\tilde d_d\simeq -7\,d_d/e \simeq -20(\bar m d_W)$ (with $\bar m \simeq 5$ MeV) 
obtained in Sect.~\ref{2HDM}, shows that the nucleon EDMs obtain contributions comparable in magnitude from all three operators, with perhaps a slight 
dominance of the Weinberg operator (although this is questionable, see the discussion below). Of course, these estimates are very rough and can in no way be used to make a definite statement about the exact sizes of the nucleon EDMs in the a2HDM scenario
that we investigate. They only provide an approximate scale for the EDMs. 

The nucleon EDM induced by the qEDM, qCEDM, and Weinberg operator has been investigated extensively in the literature (see Refs.~\cite{Pospelov_review, Engel:2013lsa} for reviews). In particular, the calculation for all three operators has been performed with QCD sum rules \cite{Pospelov_qCEDM, Pospelov_Weinberg, Pospelov_deuteron, Pospelov_review}. 
The authors of these references obtained,
   in our notation\footnote{As before, a Peccei-Quin mechanism was used to remove the $\tb$ term.}$^,$~\footnote{The difference between our $d$-quark CEDM ${\tilde d}_d$ and the one found in Refs.~\cite{Pospelov_qCEDM,Pospelov_review} is due to an explicit factor of $g_s(\Lambda_\chi)\simeq 2$ that appears in the definition of the qCEDM in these references.}$^,$~\footnote{A similar calculation was performed in Ref.~\cite{Hisano1} which found somewhat smaller coefficients for the qEDM and qCEDM. On the other hand, larger coefficients were found in Ref.~\cite{Hisano2}. These differences show
 that the uncertainties are large.},
\begin{eqnarray}\label{dn_H}
d_n = (1\pm0.5)\left( 1.4\,  d_d -0.55\,e \tilde d_d\right) \pm (0.02\,\mathrm{GeV})\, e d_W\,\,\,,
\end{eqnarray} 
with an unspecified, but significant
  error (and sign) on the coefficient of the Weinberg operator. The hierarchy between $d_d$, $\tilde d_d$, and $d_W$ then indicates that all operators contribute at the same level to the neutron EDM. The qEDM and qCEDM results are in good agreement with the chiral loop results and NDA. The result for the Weinberg operator is somewhat smaller (see the discussion in Ref.~\cite{Pospelov_Weinberg})
 but, in view of the large uncertainties involved,
 the estimates are not really in disagreement. 
The isoscalar combination $d_n + d_p$ has also been estimated 
with
 QCD sum rules~\cite{Pospelov_deuteron}, 
\begin{equation}\label{isoscalar}
d_n + d_p = (0.5\pm0.3)\, d_d - (0.2 \pm 0.1)\,e \tilde d_d  \pm (0.02\,\mathrm{GeV})\, e d_W\,\,\,,
\end{equation} 
with slightly smaller coefficients in front of the qEDM and qCEDM than 
in the case of the neutron EDM.

To summarize, in the a2HDM scenario of Sect.~\ref{2HDM}, the nucleon EDMs get contributions of roughly equal size from 
the $d$-quark EDM and CEDM  and the Weinberg operator.
 The rather large uncertainties 
in magnitude and sign of each of these  contributions
 make it impossible to 
obtain
  a firm prediction of the sizes of the nucleon EDMs. 
 We conclude that we cannot really do better than give a rough estimate of the combined contributions which sets the scale for the sizes of both the neutron and proton EDM:
\begin{equation}
 \left| d_{n,p} \right| = \Or ( e\, \tilde d_d )\,\,\,,
\label{a2HDMestimate}\end{equation}
and we do not expect the nucleon EDMs to be larger than this estimate.

When switching to the MSSM, the above relations should include the dependence on the $u$-quark (C)EDM. 
 In the most general case one may expect comparable contributions from the q(C)EDMs and the Weinberg operator to the nucleon EDMs. The above analysis of the a2HDM would then roughly hold for the MSSM as well. However, the MSSM allows for different hierarchies between the dipole operators as well, \textit{cf.} the discussion at the end of subsection \ref{suse:MSSM}. For example, in the `split SUSY' scenario of Ref.~\cite{ArkaniHamed:2004yi}
 the nucleon EDMs would be given directly in terms of the quark EDMs.

\subsection{A short intermediate summary}

Before we proceed to the discussion of light-nuclear EDMs, let us briefly summarize what we found so far.
 We have seen that the different scenarios of Sect.~\ref{3scenarios}
 predict distinct hierarchies for the $\slashPT$ pion-nucleon interactions in Eq.~\eqref{g01}. Roughly, we find $\bar g^{\tb}_1/g^{\tb}_0 \simeq -0.2$  for the $\tb$ term \cite{Jan_2013}, $\bar g^{\mathrm{LR}}_1/ \bar g^{\mathrm{LR}}_0\simeq -50$ for the mLRSM
\cite{deVries:2012ab}, and $|\bar g^{\mathrm{H}}_0|\simeq |\bar g^{\mathrm{H}}_1|$ and $|\bar g^{\mathrm{MSSM}}_0|\simeq |\bar g^{\mathrm{MSSM}}_1|$ in the a2HDM and MSSM scenarios (although $|\bar g^H_1|$ might be somewhat larger than $|\bar g^H_0|$ \cite{Pospelov_piN}).

This information, however, does not lead to a  solid prediction of
 the sizes of the neutron and proton EDM apart from the expectation that, in all scenarios, they are of comparable size. The lack of predictive power is mainly caused by the fact that the nucleon EDMs obtain leading-order contributions from tree-level diagrams independent of the $\slashPT$ pion-nucleon interactions. The situation is somewhat better for the $\tb$ term (see Ref.~\cite{Guo12}) because of lattice data,
 but the uncertainties, in particular for the proton EDM, are still significant. 
 Lattice efforts are underway to improve this situation. For the higher-dimensional BSM sources little progress is expected in the near future. 

A signal in a single EDM measurement would, of course, not point to its origin. The
  above considerations imply that, at the moment, even a measurement of both the proton and neutron EDM is not enough to disentangle the various scenarios \cite{deVries2010a}  (although a hint for the $\tb$ term might be found). Additional measurements are therefore required, and in the next sections we will argue that light-nuclear EDM experiments are excellent probes for this task. In Sect.~\ref{sec:other} we discuss EDMs of heavier systems which provide complementary information.

\section{EDMs of light nuclei}\label{lightnuclei}

The power of the $\chi$PT approach becomes much more manifest in few-nucleon systems. 
First of all, the EDMs of light nuclei can be accurately calculated in terms of the $\slashPT$ hadronic interactions. 
The associated nuclear uncertainties are much smaller than the hadronic uncertainties appearing in the LECs themselves.
Moreover, while
 the $\slashPT$ pion-nucleon vertices only contribute to the nucleon EDM at the one-loop level, 
which brings in a loop suppression and counterterms, light-nuclear EDMs depend already 
 at tree-level on the pion-nucleon vertices and counterterms only appear at subleading orders. 

In Refs.~\cite{deVries2011b, Jan_2013} EFTs\footnote{The EFTs differ somewhat in their power counting, but the leading results are identical.}
 have been constructed
with
 which controlled calculations of light-nuclear EDMs can be performed. 
In fact, this has already been done for the EDMs of the deuteron and tri-nucleon system. The calculations in Refs.~\cite{deVries2011b, Jan_2013}
 used a so-called `hybrid' approach in which the nuclear wave functions were calculated using modern phenomenological potentials 
 while the $\slashPT$ potential and currents were calculated using chiral EFT. Recently, these calculations have been repeated
  using chiral EFT for both the $PT$-even and -odd parts of the problem \cite{Jan_new,J.Bsaisou}. 
 Although, the results of the hybrid and full EFT calculations are very similar, 
 the latter approach has the advantage that the nuclear uncertainty can be quantified by varying the cut-off parameters that appear
  in the solution of the scattering equations. This gives us a quantitative handle on the nuclear uncertainties which turn out 
 to be small compared to the hadronic uncertainties in the $\slashPT$ LECs themselves. 

Before going to the actual results a few comments are in order. Based on chiral symmetry considerations, it was argued 
 in Ref.~\cite{deVries2011b} that light-nuclear EDMs should be calculated in terms of six LECs. 
So far we have only encountered four, namely the $\slashPT$ pion-nucleon LECs $\bar g_0$ and $\bar g_1$ and the nucleon 
 EDMs $d_n$ and $d_p$\footnote{In this section we will treat the EDMs $d_n$ and $d_p$ as effective parameters. The reason 
 is
 that the part of $d_{n,p}$ depending on $\bar g_{0,1}$ cannot be isolated from the tree-level LECs $\bar d_{0,1}$ in a model-independent way.}.
  The other two LECs introduced in Ref.~\cite{deVries2011b} are associated with $\slashPT$ nucleon-nucleon contact interactions of the form
\begin{equation}\label{contact}
\mathcal L_{N\!N} = \bar C_1 \Nb N\,\partial_\mu(\Nb S^\mu N) + \bar C_2 \Nb \boldtau N\cdot\partial_\mu(\Nb S^\mu  \boldtau N) \,\,\,.
\end{equation} 
For the $\tb$ term and most of the higher-dimensional BSM operators discussed above
 these contact interactions appear at high order in the chiral Lagrangians and their effects are negligible compared 
  to the one-pion exchange between
nucleons
 proportional to $\bar g_{0,1}$\cite{Mae11}. 
 However, for chiral-invariant
  sources such 
 as the Weinberg operator, the $\slashPT$ pion-nucleon LECs are suppressed and the terms in Eq.~\eqref{contact} appear 
 at the same order as $\bar g_{0,1}$. Nevertheless, the terms in Eq.~\eqref{contact} only play a marginal role as we will discuss in more detail below. 

The three-pion vertex with LEC $\bar \Delta^{\mathrm{LR}}$ appearing in Eq.~\eqref{chiralmLRSM} was not
 considered in Ref.~\cite{deVries2011b}. Although it has little consequences for the nucleon and deuteron EDMs, 
it introduces a potentially important $\slashPT$ three-body interaction \cite{deVries:2012ab} which 
 could affect the ${}^3$He and ${}^3$H EDMs. This has not been taken into account so far. 

\subsection{The EDM of the deuteron}
The EDM of the lightest bound nucleus, the deuteron, has been investigated in a number of papers in
 recent years \cite{khriplovich, Pospelov_deuteron, LiuTimmermans, Afnan, deVries2011a, deVries2011b, Jan_2013,Song:2012yh, Jan_new}. 
 From a theoretical point of view, the deuteron is particularly interesting. Not only because it is a rather simple nucleus 
 which can be accurately described, but also because its spin-isospin properties ensure that the deuteron EDM has rather distinctive properties. 

At leading order in the EFT, the deuteron EDM
 obtains two contributions. The first one
 is simply the contribution from the constituent nucleon EDMs 
 which is trivially evaluated as $d_n + d_p$. 
The second contribution
  is due to the exchange of a single pion between the nucleons 
 involving a $\slashPT$ pion-nucleon vertex (i.e., $\bar g_0$ or $\bar g_1$)
and the coupling of
  the external photon  to the proton charge.
  All calculations are consistent and here we quote the central value and uncertainty of the chiral EFT result \cite{Jan_new, J.Bsaisou}: 
\begin{equation}\label{dEDMfull}
 d_D = d_n + d_p + \bigl [(0.18 \pm 0.023)\,\bar g_1 +(0.0028\pm0.0003)\, \bar g_0\bigr ]\efm \, ,
\end{equation} 
which is almost independent of
 $\bar g_0$. This can be understood from the following reasoning: 
 The deuteron ground state is a ${}^3 S_1$ state with a small ${}^3 D_1$ admixture. After a pion exchange involving 
a leading $\bar g_0$ vertex which conserves the total isospin, the wave function obtains a small ${}^1P_1$ component. 
 Because the electric interaction with the proton charge is spin-independent, it cannot return the wave function to its 
${}^3 S_1$-${}^3D_1$ ground state and the contribution vanishes. 
 By exactly the same argument the leading contributions 
 from the $N\!N$ contact interactions in Eq.~\eqref{contact} vanish
for
  the deuteron \cite{deVries2011a}. 

The systematic nature of the EFT approach used in Refs.~\cite{deVries2011b, Jan_2013} allows the calculation of higher-order corrections, for example, due to two-pion-exchange diagrams \cite{Mae11} or two-body currents. Such corrections give rise to the small dependence of $d_D$ on $\bar g_0$. We now turn to the implications of this result for the various scenarios. 

\subsubsection{The $\tb$ scenario}
In this scenario the deuteron EDM can be given as a function of $\tb$. 
It follows from
  Eqs.~\eqref{g0theta}, \eqref{g1theta}, \eqref{isoscalartheta}, and \eqref{dEDMfull} 
that
\begin{equation}\label{dEDMtheta1}
 d_D^{\tb} =  \left[(-1.8\pm 1.4) + (0.55\pm0.36\pm0.054) -   
 (0.05\pm0.02\pm0.006)\right]\cdot 10^{-16}\,\tb\,\ecm\,\,\,,  
\end{equation}
where the first term is the contribution from the nucleon EDMs and the second and third, respectively, 
from
 the two-body contribution proportional to $\bar g_1^{\tb}$ and $\bar g_0^{\tb}$. 
 The first error of the coefficients
  is due the hadronic uncertainty in the LECs (see Eqs.~\eqref{g0theta} and \eqref{g1theta}), while the 
 second error of the last two terms is due to
  the nuclear uncertainty. The hadronic 
uncertainty is significantly larger than the nuclear uncertainty.

We can learn a few things from Eq.~\eqref{dEDMtheta1}. First of all, the deuteron is most likely dominated by the nucleon EDMs, 
although the uncertainties are still too large to say this with full confidence~\cite{deVries2011b}.
More  input from lattice calculations
 is needed to improve the situation. Second, 
when measurements of $d_n$, $d_p$, and $d_D$  will be available,
the relation 
\begin{equation}\label{thetatest_d}
d_D^{\tb}-d_n^{\tb}-d_p^{\tb} =  (5.0\pm3.7)\cdot 10^{-17}    \,\tb\,\ecm
\end{equation}
will be
  a promising and relatively precise method to 
directly
 extract the value of $\tb$ from experiments \cite{Jan_2013}. 
 The existence of the $\tb$ term can then be tested in several ways. One can compare the experimental value of $d_n$ and/or $d_p$ 
 to  lattice calculations. 
A more robust test, independent of lattice results, would be the measurement of the ${}^3$He or ${}^3$H EDM 
whose dependence on $\tb$ can be firmly predicted.
We will discuss this in Sect.~\ref{tritheta}.

\subsubsection{The mLRSM scenario}

The mLRSM scenario is in general more uncertain than the $\tb$ term. 
 Because $\bar g_0^{\mathrm{LR}}/\bar g_1^{\mathrm{LR}}\ll 1$ (see Eq.~\eqref{ratiomLRSM}) we can safely ignore
  the $\bar g_0$ term in Eq.~\eqref{dEDMfull}. 
Inserting the NDA estimates\footnote{Note that the size of $\bar g^\mathrm{LR}_1$ might be somewhat larger than the NDA estimate due 
to the large size of the term
$4c_1$, see the discussion after Eq.~\eqref{g1total}. The estimate used here is conservative. } 
given
  in Eqs.~\eqref{NDA_g1_mLRSM} and \eqref{NDA_dn_mLRSM} into the expression for the deuteron EDM, we obtain
\begin{equation}
\left| d_D^{\mathrm{LR}} \right| = \bigl | (10\,\mathrm{MeV})\,e \mathrm{Im}\Xi  \,\pm \, (100\,\mathrm{MeV})\,e \mathrm{Im}\Xi  \bigr |\,\,\,,
\end{equation}
where the first term is due to the nucleon EDMs and the second term is
 the two-body contribution. 

These rough estimates tell us that the two-body contribution to the dEDM is about an order of magnitude larger than the sum of the nucleon EDMs. Without detailed information on the LECs (for which input 
from lattice calculations
 is required) this statement cannot be made much more precise. Nevertheless, the difference between the ratios of 
 deuteron-to-nucleon-EDMs for 
the $\tb$  and the mLRSM scenario
 tell us that a deuteron EDM experiment would be  complementary to nucleon EDM experiments. In particular, a 
 large deuteron-to-nucleon EDM ratio would be indicative 
of BSM physics \cite{Pospelov_deuteron, deVries2011a}, 
 in particular  of the mLRSM scenario.

\subsubsection{The a2HDM and MSSM scenarios}
The situation for the a2HDM is somewhat similar to that of the  mLRSM.
  The $\bar g^{\mathrm{H}}_0$ term in Eq.~\eqref{dEDMfull} can be neglected since we expect $|\bar g^{\mathrm{H}}_0| \simeq |\bar g^{\mathrm{H}}_1|$, while the coefficient in front of $\bar g_0^{\mathrm{H}}$ is a hundred times smaller. 
 Using the estimates from Sects.~\ref{piN2HDM} and \ref{nEDM2HDM} gives
\begin{equation}
d_D^{\mathrm{H}} = \pm (e\,\tilde d_d) \,-\, (2 ^{+4}_{-1} )e\,\tilde d_d\,\,\,,
\end{equation}
where the first term is
 due to the nucleon EDMs\footnote{We most likely overestimate the nucleon EDM contribution to $d_D^{\mathrm{H}}$ since 
 $d^{\mathrm{H}}_n+d^{\mathrm{H}}_p$ 
 is expected to be smaller than $d_n^{\mathrm{H}}$ and $d_p^{\mathrm{H}}$ individually, 
 see Eqs.~\eqref{dn_H} and \eqref{isoscalar}. } 
 and the second term 
results from
 the two-body contribution. The uncertainty of the two-body contribution is obtained from the QCD sum rules estimate of $\bar g_1^H$~\cite{Pospelov_piN}. The nuclear uncertainty is neglected since it is at least an order of magnitude smaller. 

In the a2HDM scenario
 the deuteron-to-nucleon EDM ratio 
 lies  between the $\tb$ and mLRSM scenarios discussed above. 
 It can be expected that the deuteron EDM is a few times bigger than the sum of the nucleon EDMs. 
 However, it must be stressed that the nucleon EDMs obtain contributions from three different 
BSM operators, \textit{cf}.~Eq.~\eqref{2HDM1GeV}. Therefore,
 the accumulated uncertainty is significant. In particular the uncertainty associated with the Weinberg operator is large. 
Therefore, the conclusion $|d_D^{\mathrm{H}}|> | d_n^{\mathrm{H}}+d_p^{\mathrm{H}}|$ might be premature.

As discussed in Sect.~\ref{piN2HDM} also in the MSSM we expect $|\bar g^{\mathrm{MSSM}}_0| \simeq |\bar g^{\mathrm{MSSM}}_1|$ which means we can again neglect the $\bar g_0$ term in Eq.~\eqref{dEDMfull}. However, in the MSSM it is even harder to make a statement about the size of $d_D$ with respect to $d_{n,p}$. In case the qCEDM is significant we expect similar results as in the a2HDM. On the other hand, if the qEDMs and/or the Weinberg operator are large with respect to the qCEDM, the relation $d_D = d_n + d_p$ should hold to good approximation. More refined
statements should become possible, once the parameter space of the MSSM is further constrained.

\subsubsection{The deuteron EDM: an overview}
Let us briefly summarize the 
results on the  dEDM in  the above scenarios.
 In case of the QCD $\tb$ term, the dEDM is a relatively well understood quantity and can be 
directly
   expressed as a function of the fundamental parameter $\tb$. 
 Results so far indicate that the 
value of the dEDM is
  rather close to the sum of the nucleon EDMs, while
  the relatively small difference $d_D^{\tb}-d_n^{\tb}-d_p^{\tb}$ provides a good method to extract the value of $\tb$. 

The situation is different for the other three
 scenarios. Both for the  mLRSM and the a2HDM
  we expect 
\begin{equation}
\left |\frac{d_D-d_n-d_p}{ d_n + d_p}\right|>1\,\,\, .
\end{equation} 
However, the exact value of this ratio is uncertain. In the mLRSM, NDA suggests that the dEDM can be larger by an order of magnitude, while in the a2HDM the enhancement is more likely a factor of a few. Unfortunately, the large uncertainties involved in the LECs 
preclude
  a more quantitative statement. In the MSSM the situation is more uncertain and \mbox{$|(d_D-d_n-d_p)/(d_n+d_p)|$} can lie between zero and a factor of a few, depending on the relative sizes of the q(C)EDMs and Weinberg operator.
 This means that additional tests
are required.
  Such a test could be the following: measurements of $d_n$, $d_p$, and $d_D$ allow the extraction of $\bar g_1$ from the relation 
\begin{equation}
d_D - d_n - d_p = (0.18 \pm 0.023)\bar g_1\,\efm\,\,.
\end{equation} 
As we shall discuss in the next section, this extraction of $\bar g_1$
  allows the separation of the $\tb$ and mLRSM
 scenarios from the other two, if measurements of the EDMs of ${}^3$He and/or ${}^3$H can be made.

To conclude, it is likely that measurements of $d_n$, $d_p$, and $d_D$ would allow
to disentangle
 the $\tb$ term from the 
three
 BSM scenarios discussed here. This already 
shows  
the potential impact of the plans to measure $d_p$ and $d_D$ in storage-ring experiments.  
 Unfortunately, the three measurements are most likely not 
sufficient
  to disentangle 
the three BSM scenarios, a problem which is mainly caused by the 
poor
  information available on the hadronic $\slashPT$ LECs. 

\subsection{The EDMs of the helion and triton}\label{triEDMs}
The experimental EDM storage-ring program not only allows the possible measurement of the proton and deuteron EDMs, 
but also those of other light nuclei. In particular, measurements on the tri-nucleon EDMs are interesting from the theoretical point of view.
 These systems are simple enough 
in order to  be accurately described
  within chiral effective theory,
  with nuclear uncertainties which are small compared to the hadronic uncertainties in the LECs. 
 In addition, the tri-nucleon EDMs are complementary to the dEDM, mainly due to their much larger dependence on $\bar g_0$. 

The ${}^3$He EDM was calculated using phenomenological $PT$-even $N\!N$ potentials (including the Coulomb potential) and a 
 one-meson-exchange model for the $\slashPT$ $N\!N$ potential in Ref.~  \cite{Stetcu:2008vt} (for older work, 
 see Ref.~\cite{Avishai:1986dw}), while the no-core shell model was used to obtain the nuclear wave function.
  This framework was also applied in Ref.~\cite{deVries2011b}, 
where  the $\slashPT$ potential was  derived within chiral effective field theory, and results for the ${}^3$H EDM were also presented.
  In a  recent work, the authors of \cite{Song:2012yh} used phenomenological $PT$-invariant potentials in combination 
 with a one-meson-exchange $\slashPT$ potential, while Faddeev equations were used to solve the three-body problem. 
 The results of Refs.~\cite{Stetcu:2008vt, deVries2011b} and \cite{Song:2012yh}
on the  dependence of the tri-nucleon EDMs on the nucleon EDMs agree. However, in Ref.~\cite{Song:2012yh}
  the dependence on
$\bar g_{0,1}$ was found to be  smaller by a factor two.   This discrepancy was recently solved in Refs.~\cite{Jan_new,J.Bsaisou}
 which confirmed the smaller results of \cite{Song:2012yh} for
  $\bar g_0$ and $\bar g_1$.  
Thus, the following results apply:
\begin{eqnarray}\label{tri}
d_{{}^3\mathrm{He}} = \,\,\,\,(0.89\pm0.01)\, d_n-(0.039\pm0.01)\, d_p +\! \bigl[ (0.099\pm 0.026)\, \bar g_0 +\!(0.14\pm 0.028) \,\bar g_1 \bigr]\,\efm\,\,, \nonumber\\
d_{{}^3\mathrm{H}} =  -(0.051\pm0.01)\, d_n+(0.87\pm0.01)\, d_p -\!\bigl[ (0.098\pm 0.024)\, \bar g_0 -\!(0.14\pm 0.028) \,\bar g_1 \bigr]\,\efm\,,\, 
\end{eqnarray}with a nuclear uncertainty of the two-body contributions of approximately $25\%$. The larger uncertainty compared to the deuteron case arises from the more complicated $3$-body Faddeev calculations,
where about $20$ intermediate partial waves need to be summed in order to reach a stable result for the two-nucleon contribution~\cite{J.Bsaisou,Jan_new}. The uncertainty of the dependence on the single-nucleon EDMs is much smaller and will be neglected below.

In principle, the tri-nucleon EDMs also depend on the $\slashPT$ contact interactions of
  Eq.~\eqref{contact}. However, for the $\tb$ term \ \cite{Mae11, Jan_2013} and the mLRSM \cite{deVries:2012ab} these 
 terms only appear at next-to-next-to-leading order, and can be neglected. In the a2HDM and MSSM they are larger because of the Weinberg 
 operator \cite{deVries2011b}, but also here their contributions turn out to be small compared 
to pion-exchange contributions. We discuss in this more detail in Sect.~\ref{tri2HDM}.

Finally, the tri-nucleon EDMs could depend on the three-pion vertex proportional to $\bar \Delta^{\mathrm{LR}}$ which appears at leading order for the mLRSM (see Eq.~\eqref{chiralmLRSM}). This vertex induces a $\slashPT$ three-body interaction which, by the power-counting rules of chiral EFT, 
formally contributes at leading order \cite{deVries:2012ab}. However, this term has not been taken into account 
 in any calculation so far. It is therefore unclear whether or not it  plays an important role.

\subsubsection{The $\tb$ term}\label{tritheta}

As was the case for the dEDM, the tri-nucleon EDMs can be expressed in terms of $\tb$ with controlled uncertainty. 
We insert Eqs.~\eqref{g0theta}, \eqref{g1theta}, and \eqref{latticetheta} into Eq.~\eqref{tri} and find 
\begin{eqnarray}\label{tri_theta_EDMs}
d_{{}^3\mathrm{He}}^{\tb} &=&  \bigl [(-2.6\pm 0.80) - (1.78\pm 0.70\pm0.46) + (0.42\pm0.28\pm0.08)\bigr]\cdot 10^{-16}     \,\tb\,\ecm\,\,,\nonumber\\
d_{{}^3\mathrm{H}}^{\tb} &=&  \bigl[\ \ (1.1\pm 0.96) + (1.74\pm 0.68\pm0.44) +(0.42\pm0.28\pm0.08)\bigr]\cdot 10^{-16}     \,\tb\,\ecm\,\,,
\end{eqnarray}
where the first term in bracket denotes the nucleon EDM contribution, while the second and third 
term is, respectively, the two-body term due to $\bar g_0^{\tb}$ and $\bar g_1^{\tb}$. 
 Just as in Eq.~\eqref{dEDMtheta1} the first error is the hadronic uncertainty,
  while the second error in the two-body contributions is the nuclear uncertainty. 
 Despite the increase of the latter with respect to the deuteron case, the hadronic uncertainties are still dominant. This might change once more precise lattice results are available, see the discussion in Sect.~\ref{sec:improve}.

It is useful to combine the two-body terms into one expression:
\begin{eqnarray}
d_{{}^3\mathrm{He}}^{\tb} &=&  \bigl[(-2.6\pm 0.80) - (1.36\pm 0.88) \bigr]\cdot 10^{-16}     \,\tb\,\ecm\,\,,\nonumber\\
d_{{}^3\mathrm{H}}^{\tb} &=&  \bigl[\ \ (1.1\pm 0.96) + (2.16\pm 0.85) \bigr]\cdot 10^{-16}     \,\tb\,\ecm\,\,.
\end{eqnarray}

Several conclusions can be drawn from these relations. First of all, 
both for
  the ${}^3$He and ${}^3$H EDMs the two-body contributions add constructively to the one-body contributions. 
Second, in both cases the two-body contributions are, within the uncertainties, of similar magnitude as the one-body contributions. 
 Third, measurements of $d_n$, $d_p$, $d_D$, $d_{{}^3\mathrm{He}}$ and/or $d_{{}^3\mathrm{H}}$ allow for a relatively precise test
  of the relevance of the $\tb$ term, even without relying on any lattice results. 
 That is, the value of $\tb$ can be extracted from $(d_D -d_n - d_p)$, which can then be compared with the predictions  
\begin{eqnarray}\label{thetatest_3H}
d_{{}^3\mathrm{He}} - 0.89\, d_n + 0.039\, d_p &=& - (1.36\pm 0.88)\cdot 10^{-16}     \,\tb\,\ecm\,\,,\nonumber\\
d_{{}^3\mathrm{H}} +0.051\, d_n -0.87\, d_p &=&\ \  (2.16\pm 0.85)\cdot 10^{-16}     \,\tb\,\ecm\,\,.
\end{eqnarray}
Of course, using lattice data would allow for additional nontrivial tests.

\subsubsection{The mLRSM scenario}\label{trimLRSM}

Within the mLRSM, the analysis of the tri-nucleon EDMs is very similar to that of the dEDM. 
 Because of the smallness of $\bar g_0^{\mathrm{LR}}/\bar g_1^{\mathrm{LR}}$, the terms 
proportional to $\bar g_0$ in Eq.~\eqref{tri} can be neglected. 
 The estimates in Eqs.~\eqref{NDA_g1_mLRSM} and \eqref{NDA_dn_mLRSM} then tell us that the tri-nucleon EDMs are,
just as the dEDM,
 about an order of magnitude larger than the nucleon EDMs. 
In particular, 
assuming
 that the nucleon EDM contribution can be neglected at leading order, the mLRSM predicts 
\begin{equation}
d_{{}^3\mathrm{He}}^{\mathrm{LR}} \simeq d_{{}^3\mathrm{H}}^{\mathrm{LR}} \simeq 0.7\, d_{D}^{\mathrm{LR}}\,\, .
\end{equation}
That is, in this scenario these dipole moments
  have the same sign and are of the same order of magnitude.

Two caveats exist that could alter this prediction. First of all, the ratio of the two-body-to-one-body contributions 
has been estimated by NDA. It is not impossible that the nucleon EDM contributions are more important than NDA suggests. 
A better test then would be to extract $\bar g_1$ from $(d_D- d_n - d_p)$ and use this to predict 
$(d_{{}^3\mathrm{He}} - 0.89 d_n + 0.039 d_p)$ and/or  $(d_{{}^3\mathrm{H}} +0.051 d_n -0.87 d_p)$. 
However, even this prediction might be altered by the second caveat which consists of possible contributions to the 
 tri-nucleon EDMs proportional to the three-pion vertex $\bar \Delta^{\mathrm{LR}}$ in Eq.~\eqref{chiralmLRSM}. 
 If these contributions are significant, both tests described above will fail because the tri-nucleon EDMs 
 depend on an independent LEC which does not appear in the leading-order expressions 
 of the nucleon and deuteron EDMs\footnote{It should be noted that $\bar \Delta^{\mathrm{LR}}$ is expected 
 to contribute to the dEDM at next-to-leading order \cite{deVries:2012ab}. 
 However, its precise contribution has not been calculated so far.}. 
We conclude that a calculation of the dependence on the tri-nucleon EDMs on $\bar \Delta^{\mathrm{LR}}$ is an important 
open
 problem. 

\subsubsection{The a2HDM and MSSM scenarios}\label{tri2HDM}
The analysis of the tri-nucleon EDMs 
within the
  a2HDM is more complicated than in the previous two scenarios. 
 Similar to the dEDM, the tri-nucleon EDMs are most likely larger than the nucleon EDMs by a factor of a few. 
 However, the exact size of $d^{\mathrm{H}}_{{}^3\mathrm{He}, {}^3\mathrm{H}}/d^{\mathrm{H}}_{n,p}$ is uncertain. 

In addition, even with measurements of $d_n$, $ d_p$, and $d_D$, the tri-nucleon EDMs cannot be firmly predicted. 
 This can be understood from the $\bar g_0$ terms in Eq.~\eqref{tri} which are expected to be significant in 
 the a2HDM scenario, but the size of $\bar g_0$ cannot be obtained from $ d_n$, $ d_p$, and $d_D$. 
 One could then think of a negative test: measurements of $ d_n$, $ d_p$, and $d_D$ allow the extraction of $\bar g_1$. 
 This value, in combination with $d_n$ and $d_p$, can be used to predict the tri-nucleon EDMs. 
 If these predictions 
would  not agree with the data, it would indicate
 that the tri-nucleon EDMs obtain an independent contribution, suggesting that $\bar g_0$ 
 plays a role which would 
hint at
 the a2HDM. 
 A caveat is that such a scenario could also point 
to the mLRSM  in which the independent contribution is due to $\bar \Delta^{\mathrm{LR}}$.

A better method to test the a2HDM scenario then seems to be the following: from measurements of $ d_n$, $ d_p$, and $d_D$, 
it is possible to extract the size of $\bar g^{\mathrm{H}}_1$. This value, in combination with a measurement of $d_{{}^3\mathrm{He}}$ ($d_{{}^3\mathrm{H}}$), allows for the extraction of $\bar g^{\mathrm{H}}_0$. 
The value of  $d_{{}^3\mathrm{H}}$ ($d_{{}^3\mathrm{He}}$) can then be predicted. 

Lattice calculations could improve this somewhat bleak scenario where five EDMs are necessary for a proper test. 
Because in the a2HDM,  the nucleon EDMs depend on the 
 EDM and CEDM of the $d$ quark  and on the Weinberg operator,
  lattice calculations of the nucleon EDM will be very difficult.
  On the other hand, the $\slashPT$ pion-nucleon LECs mainly depend on the qCEDM. 
 If $\bar g_{0,1}$ can be calculated as a function of the qCEDM, the number of necessary experiments can be reduced. 
 
If in case of the MSSM the qCEDM turns out to be significant the pattern of tri-nucleon EDMs would be similar to that of the a2HDM. 
 That is, the tri-nucleon EDMs are expected to depend on $\bar g_0$ as well. However, if the qEDM and/or Weinberg operator dominate the qCEDM, $d_{{}^3\mathrm{He}}$ ($d_{{}^3\mathrm{H}}$) are expected to lie close to $d_n$ ($d_p$). 

Finally we comment on the $\slashPT$ contact LECs in Eq.~\eqref{contact}. As discussed in 
  Refs.~\cite{Mae11, deVries2011b}, for most $\slashPT$ dimension-four and -six operators these terms are very small.
  For the Weinberg operator, however, which appears in the a2HDM and MSSM scenarios, these operators could be as important 
as one-pion exchange between nucleons involving $\bar g_0$. As we argued in Sect.~\ref{piN2HDM}, in the a2HDM the 
 contribution from the Weinberg operator to $\bar g_0$ can be neglected, because of the larger contribution 
 from the down-quark CEDM. This implies that the contributions from the interactions in Eq.~\eqref{contact} 
 to the $\slashPT$ $N\!N$ potential can be neglected as well. 
 In addition, Ref.~\cite{deVries2011b} found that the dependence of the tri-nucleon 
 EDMs on $\bar C_{1,2}$ was smaller then expected by power counting. This last point could imply that also in the MSSM
 it is safe to neglect the $\slashPT$ nucleon-nucleon contact interactions in the tri-nucleon EDMs. However, the contact interactions might become more important in heavier nuclei.

\subsection{Tri-nucleon EDMs: an overview}
For probing
  the QCD $\tb$ term the tri-nucleon EDMs are very promising observables. 
 Because the tri-nucleon EDMs depend on $\bar g_0$ at leading order, the two-body contributions are a few times bigger than for the deuteron EDM which implies that the tri-nucleon EDMs are not dominated by the constituent nucleon EDMs. The EDMs of the helion and trition are thus expected to be larger than the EDMs of the neutron and proton.
Furthermore, the small nuclear uncertainties allow for a proper test of strong $\CP$ violation, once $\tb$ has been determined from measurements of $d_n$, $d_p$, and $d_D$, or from lattice calculations in combination with a measurement of $d_n$ and/or $d_p$. 

Measurements of the tri-nucleon EDMs would 
 also provide important information on the mLRSM scenario.
  In particular, the $\slashPT$ two-body interactions dominate over the nucleon EDMs by an order of magnitude,
as  in the  case of the deuteron EDM.
       If this were
  the whole story, this would imply that $d_D$, $d_{{}^3\mathrm{H}}$, and $d_{{}^3\mathrm{He}}$ 
 depend only on a single LEC $\bar g_1^{\mathrm{LR}}$ which means that once one of these EDMs has been measured, 
 the other two can be predicted. However, the tri-nucleon EDMs might obtain an important contribution from 
the three-pion vertex proportional to $\bar \Delta^{\mathrm{LR}}$. A more conclusive statement 
 can be made once the dependence of the tri-nucleon EDMs on $\bar \Delta^{\mathrm{LR}}$ has been calculated. 

In the a2HDM scenario, the tri-nucleon EDMs are in principle independent from the 
 deuteron EDM because of the dependence on $\bar g_0^H$. Estimates of the nucleon EDMs and the 
 pion-nucleon LECs $\bar g_{0,1}^{\mathrm{H}}$ suggest that the two-body contributions dominate the light-nuclear EDMs --
  however, the uncertainties are large. 
 A lattice calculation of the $\bar g_{0,1}^{\mathrm{H}}$ induced by the qCEDM could significantly improve the situation. A lattice calculation of $d_{n,p}$ would be beneficial as well, but more complicated because of 
its dependence on the three  BSM operators in Eq. \eqref{2HDM1GeV}.

Depending on the hierarchy between the BSM operators in Eq.~\eqref{SUSY1GeV}, the situation in the MSSM might be very close to the a2HDM. This implies that these scenarios cannot be disentangled using light-nuclear EDMs alone. On the other hand, if  in the MSSM the qEDMs or the Weinberg operator are significantly larger than the qCEDMs, this would imply that $\slashPT$ two-body effects in light-nuclear EDMs are relatively small compared to contributions from the nucleon EDMs. In this particular, and admittedly \textit{ad hoc}, case, the EDMs of the deuteron and tri-nucleon EDMs should be well approximated by their constituent nucleon EDMs. Thus, the MSSM might leave a footprint behind in the hierarchy of light-nuclear EDMs which is distinct from the other scenarios. 

In any case, measurements of the tri-nucleon EDMs would provide important information on the 
 source of non-KM $\CP$ violation, if such a source exists.
 Measurements of the nucleon, deuteron, and tri-nucleon EDMs allow one
 to disentangle the $\tb$ and mLRSM scenarios from the a2HDM and MSSM scenarios considered in this paper. It is in general not possible to separate the latter two from each other using light-nuclear EDM measurements alone.

\section{EDMs of other systems}\label{sec:other}
In this section we briefly discuss EDMs of other systems which are not the main focus of this work. 
In particular 
we consider, within the above scenarios, $\slashPT$ effects in the
paramagnetic 
 atom/molecules ${}^{205}$Tl, YbF, and ThO, which depend on the electron EDM (eEDM) and semi-leptonic four-fermion operators.
We also discuss the EDM of the diamagnetic  ${}^{199}$Hg atom. There exist strong experimental limits on these EDMs,
 but atomic and nuclear theory is required to relate 
 the existing experimental bounds on $T$-violating effects in these
 complicated systems to  an  underlying mechanism of $\CP$ violation.
\subsection{The EDMs of paramagnetic systems}
\label{sec:eEDM}
So far we have focused on hadronic EDMs, but the electron EDM is, of course, an 
important
 observable as well. In general, eEDM measurements are complementary to hadronic EDM measurements because they probe 
 different fundamental parameters. The eEDM, however, is not measured directly but
 inferred from measurements on atomic and molecular systems. The current strongest bound on 
 the eEDM comes from the limit on 
a $T$-violating effect in
 the molecule ThO \cite{Baron:2013eja}.
Strong limits are obtained from the molecule YbF \cite{Hudson:2011zz,Kara:2012ay}
and the paramagnetic atom ${}^{205}$Tl \cite{Regan:2002ta} as well. 
 The eEDM is not the only $\slashPT$ source 
that would generate an  EDM of   ${}^{205}$Tl or the  $T$-violating effects searched for in
   YbF and   ThO. In particular, additional contributions can
     arise from $\slashPT$ semi-leptonic four-fermion operators,
      but, as discussed below, these  contributions can be neglected
      as compared to the one due to the electron EDM.

In the  $\tb$ scenario, that is to say, in the SM with massless
  neutrinos and a nonzero  $\tb$ term,
  the eEDM is generated as a spill-over from the quark sector by the  $\tb$ term and the KM phase
 and is therefore  much smaller than  the EDM of a
 nucleon~\cite{Choi:1990cn,Hoogeveen:1990cb,Pospelov:1991zt},
  $|d_e|\lesssim 10^{-37}\,e$ cm.
 Contributions from  $\slashPT$ semi-leptonic interactions to $T$-violating effects in atoms and molecules
might be larger than those from the eEDM, but also they are strongly suppressed~\cite{Fischler:1992ha}. 
 Therefore, in the $\tb$ scenario we do not expect a nonzero
 measurement of a $T$-violating effect in the above paramagnetic systems.

In the mLRSM, both the qEDMs and the eEDM are generated at
one-loop. These expressions  differ because the diagram for the qEDM
involves quarks 
 whereas in the case of the eEDM the loop involves massive neutrinos.
 The  qEDMs $d_u$ and $d_d$ involve the factor $\sum_{i=d,s,b}\text{Im}(e^{-i\al}m_i V_{L}^{ui}V_{R}^{ui*})$
 and $\sum_{j=u,c,t}\text{Im}(e^{-i\al}m_j V_{L}^{jd}V_{R}^{jd*})$,
 respectively  \cite{Beall:1981zq,Ecker:1983dj,Xu:2009nt}, while the expression 
  for the electron EDM contains the factor
  $\text{Im}(e^{-i\al}(M_{\nu_D})_{ee})$ \cite{Nieves:1986uk,
    Chen:2006bv}, where $M_{\nu_D}$ is the neutrino Dirac-mass matrix.
 Because the expressions for $d_q$ and $d_e$ involve different
 parameters their relative magnitudes cannot be reliably compared in
 general. However, if we assume the $\CP$ phases in both cases to be
 of the same order and take the $ee$ element of the neutrino
 Dirac-mass matrix to be of the of order the electron mass,
 $|M_{\nu_D}|\simeq m_e$, the eEDM will be suppressed 
 with respect to the qEDM
  by at least a factor $m_e/m_{u,d}$. 
 Since, in the mLRSM, the qEDM  make negligible contributions
 \cite{Xu:2009nt} to
 hadronic and nuclear EDMs as compared to the dominant contribution of
 the tree-level generated four-quark operator discussed in Sect.~\ref{mLRSM}, 
 the eEDM is expected to be significantly smaller than the nEDM. Assuming $|M_{\nu_D}|\simeq m_e$ and the different phases to be of the same order, we estimate $d_e/d_n\sim 10^{-4}$. We emphasize that a more precise statement is not possible because the
 hadronic and electron EDMs depend on different parameters.

In the version of the a2HDM 
  discussed in  Sect.~\ref{2HDM} the contribution to the eEDM is
  dominated by two-loop diagrams and the expressions are nearly
  identical to those of the $d$-quark EDM. Two things are
  altered. First of all, there is the obvious difference between the masses and
  charges of the $d$ quark and the electron. More important is that the eEDM depends on a
  different $\CP$-odd parameter, namely,
  $\text{Im}(\varsigma_l\varsigma_u^*)$ where $\varsigma_l$ is defined
  analogously to $\varsigma_d$~\cite{Pich:2009sp}, \textit{cf.} Eq.~\eqref{eq:speca2hdm}. This means it will
  be hard to compare the $d$-quark EDM and eEDM in general.
If we assume the two $\CP$-violating parameters to be of the same order, the magnitudes of
the two EDMs should be comparable at the electroweak scale with a minor enhancement of the $d$-quark EDM by a factor $m_d/m_e$.
However, the $d$-quark EDM gets large contributions from the $d$-quark CEDM when the operators are evolved to lower energies. 
It is more interesting to look at the electron-to-neutron-EDM ratio in the a2HDM. Assuming the $\CP$-violating parameters to be equal (see Eq.~\eqref{ap:relqlp}) and using Eq.~\eqref{a2HDMestimate} we estimate $|d_e/d_n|\sim 10^{-2}$. The upper bound on the eEDM then implies a bound on the nEDM $d_n\lesssim 10^{-26}\,e\, \text{cm}$ in this scenario. In view of the dependence of the eEDM and nEDM on different parameters this bound is not very stringent.

 In the MSSM one expects, in general, the
    electron-to-neutron-EDM ratio to be of the same order
        of magnitude as in the a2HDM. In particular, the `split SUSY'
     scenario predicts a strong correlation between these EDMs,
     $ |d_e/d_n| \sim 1/10$  \cite{Giudice:2005rz}, up to theoretical
uncertainties in the calculation of $d_n$.

What about the semi-leptonic operators in the BSM scenarios? In both the mLRSM and the a2HDM they are generated through tree-level exchange of a heavy Higgs boson. 
These operators are therefore
   suppressed by a factor $m_q m_e/v^2$  from the Yukawa couplings. In addition there is a factor $1/m_H^2$ from Higgs-boson exchange
$H$, where $m_H$ denotes the mass of $H$.
   While in the a2HDM $m_H$ must not exceed $\sim 1$ TeV, in the mLRSM
   the masses of the heavy Higgs bosons  with $\CP$-violating
      Yukawa couplings are of the order of $10$ TeV.
      
Even though in both models the eEDM is
            generated at the loop level, it nevertheless  dominates the
contributions   to the above atomic/molecular $T$-odd effects. The only
way around this  is if the $\CP$ phases that appear in $d_e$ are tuned to be much smaller than the phases of the  coefficients of the
semi-leptonic operators. Barring this possibility then, in the a2HDM,
the contributions  from the semi-leptonic operators
          to the  $T$-violating effects in   $^{205}$Tl, ThO,
        and YbF are suppressed by about two orders of magnitude with
       respect to those of the eEDM, see Ref.~\cite{Jung:2013hka} for a
more detailed discussion. In the mLRSM the contributions of the
${\slashPT}$ semi-leptonic operators to these paramagnetic systems are
even less important  than in the a2HDM, in view of the larger
suppression factors discussed above.
       In the MSSM,  the ${\slashPT}$ semi-leptonic operators are
non-negligible, especially when the first and second
         generation of sfermions are very heavy \cite{Lebedev:2002ne}.
However, because global fits to experimental data
          seem to disfavor large values of $\tan\beta \gtrsim 30$ \mbox{(\textit{cf}.
Sect.~\ref{suse:MSSM})}, the dominant contribution to,
         e.g., the $^{205}$Tl EDM still comes from the electron EDM
\cite{Lebedev:2002ne}. Thus one may conclude that  the eEDM provides the
dominant contribution to the $T$-violating effects in  $^{205}$Tl, ThO,
and YbF.

In summary, the size of the eEDM with respect to hadronic EDMs gives
additional information to disentangle the various scenarios. Clearly, a
nonzero eEDM would rule out the pure $\bar\theta$-scenario.
 Within the  a2HDM and the MSSM, the eEDM is expected to be about one to
two orders of magnitude smaller  than  the neutron
EDM. In the mLRSM this suppression is expected to be even larger.
However, in these scenarios no solid predictions can be made because the
eEDM and the hadronic EDMs depend, in general, on different unknown
  parameters.

\subsection{The ${}^{199}$Hg EDM}

Schiff's theorem \cite{Schiff:1963zz} ensures that in the non-relativistic limit the EDM of a point-like nucleus in an atomic system is completely screened by the electrons surrounding the nucleus. This would imply that the total EDM of an atomic system is zero. However, in real atoms the necessary conditions for Schiff's theorem to apply are violated. For example, in 
case of ${}^{199}$Hg, 
 a diamagnetic atom, the largest contribution to the atomic EDM stems from the finite size of the nucleus and is induced by the so-called nuclear Schiff moment $S_{\mathrm Hg}$\footnote{The mercury EDM also receives contributions from the electron EDM and $\slashPT$ semi-leptonic interactions, but these are better probed in the paramagnetic systems discussed in the previous section. We therefore do not discuss the (semi-)leptonic contributions here.}. For ${}^{199}$Hg, the relation between the atomic EDM, $d_{\mathrm{Hg}}$, and $S_{\mathrm{Hg}}$ is given by~\cite{Engel:2013lsa, Dzuba:2002kg, Dzuba:2009kn, Dzuba:2012bh}
\begin{equation}\label{atom}
d_{\mathrm{Hg}}= (2.8\pm0.6)\cdot 10^{-4}\, S_{\mathrm{Hg}}\,\mathrm{fm}^{-2}\,\,\,,
\end{equation}
with an uncertainty estimate based on  Ref.~\cite{Dmitriev:2003kb}. While the atomic calculation is rather well under control,
the main uncertainties arise from the nuclear-theory calculation of $S_{\mathrm{Hg}}$. Typically it is calculated as a function of the pion-nucleon couplings, {\it cf.}~Eq.~\eqref{g01}, and the single nucleon EDMs. However, at present there exists no EFT for nuclei with this many nucleons. It is therefore not clear whether or not there will be important contributions from other $\slashPT$ hadronic interactions such as the contact interactions in Eq.~\eqref{contact}. In addition, corrections to leading terms cannot be systematically calculated which means that the uncertainties are difficult to quantify. If we assume that $S_{\mathrm{Hg}}$ is dominated by pion-nucleon interactions, the estimated uncertainties are large \cite{deJesus:2005nb, Ban:2010ea, Engel:2013lsa}
\begin{equation}
S_{\mathrm{Hg}} = \left[(0.37\pm0.3)\bar g_0 + (0.40 \pm 0.8)\bar g_1\right]\,e\,\mathrm{fm}^3 \,\,\,.
\label{shg}
\end{equation} 
For example, in case of the $\tb$ scenario we can use Eqs.~\eqref{g0theta} and \eqref{g1theta} to obtain\footnote{In case of the $\tb$ term, $S_{\mathrm{Hg}}^{\tb}$ receives contributions from the nucleon EDMs which are of the same order as the $\bar g_{0,1}^{\tb}$ contributions \cite{Dmitriev:2003kb}. We do not give the detailed expressions here. }
\begin{equation}\label{Schifftheta}
S_{\mathrm{Hg}}^{\tb} = \left[-(6.5\pm2.5\pm5.3) + (1.2 \pm 0.8\pm2.4)\right]\cdot 10^{-3}\,\tb\,e\,\mathrm{fm}^3 \,\,\,,
\end{equation} 
with the first term due to $\bar g^{\tb}_0$ and the second due to $\bar g_1^{\tb}$. In each bracket the first error is the hadronic uncertainty
from the coupling constants and the second the nuclear uncertainty taken from Eq.~\eqref{shg}. 
In contrast to the results for the  EDMs
of light nuclei, here the nuclear uncertainty is dominant and might be difficult to reduce. Inserting Eq.~\eqref{Schifftheta} into Eq.~\eqref{atom} and combining all uncertainties gives
\begin{equation}
d_{\mathrm{Hg}}^{\tb}= -\left(1.5\pm1.8\right)\cdot 10^{-19}\,\tb\,e\,\mathrm{cm}\,\,\,,
\end{equation}
to which the contributions from the constituent nucleon EDMs still need to be added.
This result implies that even if a nonzero EDM was measured for $^{199}$Hg,
the uncertainties, at the moment, would be too large to test the $\tb$ scenario.

In the mLRSM scenario, the dominant contribution to $S_{\mathrm{Hg}}$ is expected to come from $\bar g_1^{\mathrm{LR}}$. However, due to the large nuclear uncertainty it is not possible to predict the size of $d_{\mathrm{Hg}}$ once $\bar g_1^{\mathrm{LR}}$ has been extracted from, for example, light-nuclear EDM experiments. For the same reason a measurement of $d_{\mathrm{Hg}}$ cannot be used to extract a
sufficiently precise value of $\bar g_1^{\mathrm{LR}}$. The discussion for the a2HDM and MSSM\footnote{For a recent analysis of Schiff moments
 within the MSSM, see Ref.~\cite{Ellis:2011hp}.} scenarios is similar to the mLRSM scenario. In these cases, also $\bar g_0^{\mathrm{H}}$ is expected to give a significant contribution, but again the nuclear uncertainties are too large to 
extract any nontrivial, quantitative information.  In conclusion, disentangling the various scenarios using measurements of 
$d_{\mathrm{Hg}}$ 
is not possible, unless the nuclear theory is improved substantially\footnote{Future experiments on ${}^{225}$Ra might be more promising since the nuclear theory is more reliable than for ${}^{199}$Hg \cite{Engel:2013lsa}. }. 

\section{Discussion, outlook, and summary}\label{disc}
\subsection{Testing strategies}\label{sec:test}
Based on the findings of this paper
there are several strategies to reveal nontrivial information on the $\slashPT$ sources, once non-vanishing measurements and/or improved lattice calculations of EDMs of nucleons and light nuclei are available\footnote{In this section we do not consider measurements of the EDM of ${}^3$H since, due to its radioactive nature, it is not likely to be measured in a storage ring experiment.}.

Since the $\slashPT$ pion-nucleon coupling constants are known quantitatively for a non-vanishing $\bar{\theta}$-term, the most conclusive tests can
be formulated for this scenario: if the QCD $\tb$-term is assumed to be the prominent source of $\CP$ violation beyond the CKM-matrix,
it follows directly from  Eq.~(\ref{tri_theta_EDMs}) that the value of  $\bar{\theta}$ can be extracted from EDM measurements of both the
neutron and $^3$He via
\begin{equation}
d_{{}^3\mathrm{He}} - 0.9\, d_n = (-1.4 \pm 0.9)\cdot 10^{-16}     \,\tb\,\ecm \ ,
\label{thetatest1}
\end{equation}
where the uncertainties were added in quadrature and we dropped the contribution from the proton EDM, whose contribution to  $d_{{}^3\mathrm{He}}$
is  strongly suppressed. A lattice calculation for $d_n$ then allows for the first nontrivial test of the assumed scenario.
Note that in this case all nonperturbative QCD effects can be controlled quantitatively. In the next subsection we
discuss how the present uncertainty of about 70\% can be reduced further. 
The value of $\bar{\theta}$ extracted from Eq.~\eqref{thetatest1} can now be used to predict 
\begin{equation}
d_{D} - \, d_n - \, d_p = (5 \pm 4)\cdot 10^{-17}     \,\tb\,\ecm \ ,
\label{thetatest3}
\end{equation}
where the uncertainties displayed in Eq.~\eqref{dEDMtheta1}  were again added in quadrature. This provides the second
nontrivial test, if in addition also the EDM of the deuteron, $d_{D}$, and of the proton, $d_p$, were measured.
If  the same value of  $\bar{\theta}$ could explain simultaneously the measured values of $d_{{}^3\mathrm{He}}$, $d_D$ as well as
$d_n$ and $d_p$ calculated on the lattice, it would provide very strong evidence that indeed the QCD $\tb$-term is the
origin of the non-vanishing EDMs.

If the QCD $\tb$-term would not pass this test, alternative scenarios need to be studied. In this work we considered, for illustration, the mLRSM, the a2HDM, and the MSSM. In these cases the absence of a quantitative knowledge of the induced LECs
hinders predictions for the nuclear EDMs analogous to Eqs.~\eqref{thetatest1} and \eqref{thetatest3}. However, 
at least for the mLRSM different EDMs can be related to each other, since in this scenario $\bar g_1$ dominates
over $\bar g_0$. This results in the prediction that the single nucleon EDM should be significantly smaller
than the nuclear ones. In addition, one may extract $\bar g_1^{\rm LR}$ from Eq.~\eqref{dEDMfull}
$$
d_D - (d_n + d_p) \simeq d_D =  (0.18 \pm 0.02)\,\bar g_1^{\rm LR} \efm \ .
$$
This value of $\bar g_1^{\rm LR}$ can then be used to predict $d_{{}^3\mathrm{He}}$ according to Eq.~\eqref{tri}
$$
d_{{}^3\mathrm{He}} - 0.9\, d_n  \simeq d_{{}^3\mathrm{He}} = (0.14\pm 0.03) \,\bar g_1^{\rm LR} \,\efm \ .
$$
Note, this nice relation could be spoiled by a potentially large $\CP$-odd three-body force, as discussed above.
If the mLRSM fails its test, too, the physics responsible for the $\CP$-violation must come from yet another
theory beyond the Standard Model, candidates being the a2HDM and the MSSM discussed in this work: in these
models $\bar g_0$ and  $\bar g_1$ are expected  to be similar in size and the single-nucleon as well as
$^3$He EDMs acquire additional important contributions.
As outlined in Sect.~\ref{sec:eEDM} the electron EDM might provide additional information to disentangle
the mLRSM from the other scenarios.

Thus, if the EDMs for proton, neutron, deuteron, and $^3$He were measured with high precision, highly nontrivial information could
be deduced on the $\CP$-violating physics responsible for their appearance.

\subsection{Expected sensitivities}\label{sec:bounds}
\begin{figure}[t]
  \centering
 \includegraphics[width=0.7\textwidth]{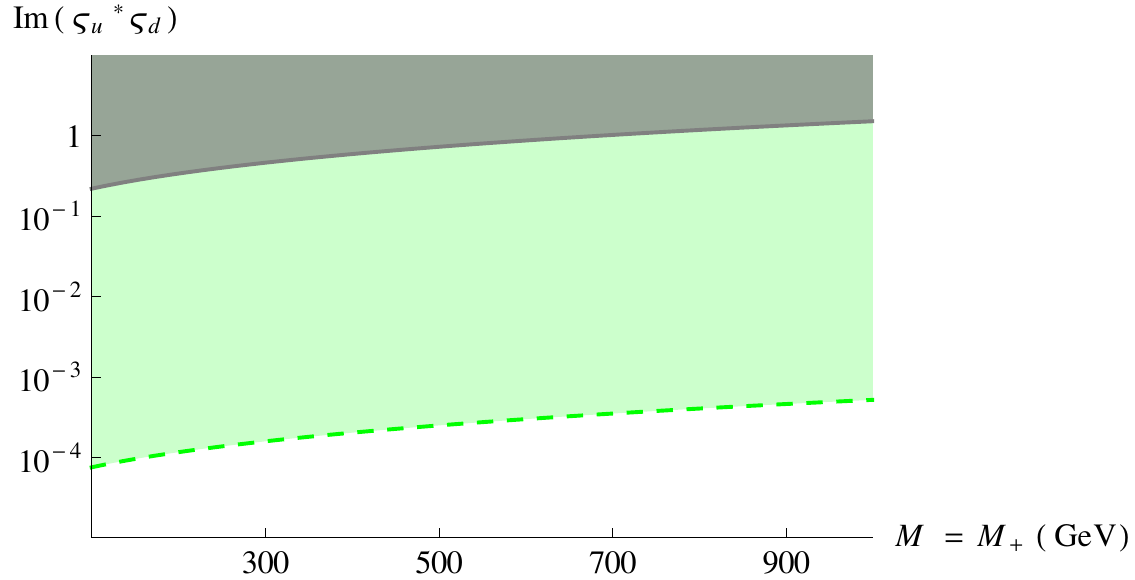}   
\caption{\footnotesize Sensitivities to the magnitude of the $\CP$-violating parameter $\mathrm{Im}(\varsigma_u^*\varsigma_d)$ as a function of the mass of the additional Higgs fields. The region of parameter space which is excluded by the current upper limit on the neutron EDM ($d_n \leq 2.9\cdot 10^{-26}\,e \,\text{cm}$) is shown in grey and bounded by the solid line. The region that would be probed by a measurement of the proton EDM at the accuracy of $10^{-29}$ e cm is shown in green and bounded by the dashed line. }
 \label{a2HDMbounds}
\end{figure}

So far we have focused on ways to disentangle the four scenarios of $\CP$ violation. However, it is also interesting to see how well the current and proposed EDM experiments are able to probe non-KM $\CP$ violation in each of the four scenarios. To this end we will discuss the sensitivities to the $\CP$-violation parameters that would result from EDM measurements of the proton, deuteron, and helion at the envisaged accuracy \cite{Farley:2003wt,Onderwater:2011zz,Pretz:2013us,JEDI}. These sensitivities are shown in Table \ref{bounds} for the $\tb$ and mLRSM scenarios and in Fig.~\ref{a2HDMbounds} for the a2HDM scenario. We do not discuss the MSSM scenario here. For comparison, we also show the bounds that can be set by the current upper limit on the neutron EDM \cite{Baker:2006ts}. The most conservative values allowed by the uncertainties of the expressions in Sects.~\ref{chiral} and \ref{lightnuclei} were used to obtain these bounds and sensitivities.
We assigned an uncertainty of a factor $10$ to the estimates based on NDA.

\begin{table}[t]

\begin{center}\footnotesize
\begin{tabular}{||c|c|ccc||}
\hline
    \rule{0pt}{2.5ex}&$d_n$ & $d_p$ & $d_D$ &$d_{{}^3\mathrm{He}}$\\
 \hline
$\tb$ & $  1\cdot 10^{-10}$ &$ \times$ & $\times$&
  $ 4\cdot 10^{-14} $\\  
 $\sin\zeta \,\mathrm{Im}(V_L^{ud*}V_R^{ud}e^{i\al})$&  $  4\cdot 10^{-5} $& $ 2\cdot 10^{-8}$& $  2\cdot 10^{-9}$ & $  2\cdot 10^{-9}$\\
 \hline
\end{tabular}
\end{center}
\caption{\small Sensitivities to the magnitudes of $\CP$-violating parameters of the $\tb$ and mLRSM scenarios. The first column shows the relevant parameters, while the second column shows bounds from the current upper limit on the neutron EDM, $d_n \leq 2.9\cdot 10^{-26}e \,\text{cm}$. The remaining columns show the values to which the $\CP$-violating parameters could be probed by measurements of the proton, deuteron, and helion EDMs at the envisaged accuracy: $ 10^{-29}$ e cm.  The contributions of $\tb$ to the proton and deuteron EDMs are consistent with zero within the uncertainties of Eqs.~\eqref{latticetheta} and \eqref{dEDMtheta1} and can therefore, at the moment, not be used to probe $\tb$. If we were to take the central values of Eqs.~\eqref{latticetheta} and \eqref{dEDMtheta1}, we would obtain sensitivities of $9\cdot 10^{-14}$ and $8\cdot 10^{-14}$ for $d_p$ and $d_D$, respectively.}\label{bounds} 
\end{table}

The current upper limit on the neutron EDM already stringently constrains the $\CP$-violating parameters appearing in each of the three scenarios\footnote{The bound on the $\CP$-violating parameters of the mLRSM in Table \ref{bounds} would be an order of magnitude stronger if we would not include the factor $10$ uncertainty assigned to the NDA estimate. Such a bound would still be roughly an order of magnitude weaker than the bound derived in Ref.~\cite{Zhang:2007da}, see Ref.~\cite{Seng14} for more details.}. Obviously, for all scenarios a measurement of the proton EDM at the proposed accuracy would greatly improve the sensitivity to the $\CP$-violating parameters as compared to the current neutron EDM limit.

To what extent the deuteron and/or helion EDMs are more sensitive than the proton EDM depends on the relative sizes of the corresponding EDMs which differ between the various scenarios. In the $\tb$ scenario, the helion EDM should be a few times bigger than the deuteron and proton EDMs, while in the mLRSM scenario both the deuteron and helion EDMs are expected to be an order of magnitude larger than the proton EDM. This is reflected in Table \ref{bounds}, where the greatest sensitivity to $\tb$ would come from a helion EDM measurement, while a deuteron EDM measurement would be the best probe for the mLRSM scenario. In the a2HDM scenario, the deuteron and helion EDMs are expected to be of similar size, both are larger than the proton EDM by a factor of a few. However, the exact size of this factor is rather uncertain and we therefore only show in Fig.~\ref{a2HDMbounds}
the region of parameter space which could be probed by a measurement of the proton EDM at an accuracy of $10^{-29}e \,\text{cm}$.

\subsection{How to improve the theoretical accuracy}\label{sec:improve}
Various paths are possible to improve the theoretical accuracy of the EDM calculations
and make,  in this way,  the tests outlined in Sect.~\ref{sec:test} even more challenging for the models.

Let us first consider $\CP$ violation due to the QCD $\tb$-term.
An improved theoretical understanding of single-nucleon EDMs can come from
lattice QCD only. Respective calculations near or even at the physical pion mass including the
still missing estimates of the pertinent systematic errors are necessary. In this way 
the parameter $\bar\theta$ could be determined from a measurement of, e.g.,
the neutron EDM and used to predict the EDMs of the proton and the light nuclei 
based on Eqs.~\eqref{thetatest1} and \eqref{thetatest3}.

The uncertainties of the two-nucleon contributions are dominated by those of the coupling constants, i.e., the
nuclear part of the calculations is sufficiently well under control.  This is clearly visible in the deuteron result Eq.~\eqref{dEDMtheta1}, 
where the nuclear uncertainty is only about $15$\% of the hadronic one.
For the helion and triton calculations the nuclear uncertainty  increases to about
$60$\% of the hadronic one, {\it cf.} Eq.~(\ref{tri_theta_EDMs}). 

To reduce the uncertainties of the $\slashPT$ pion-nucleon
coupling constants $\bar g_0^\theta$ and $\bar g_1^\theta$, {\it cf.}  Eqs.~(\ref{g0theta}) and (\ref{g1theta}), respectively,
again lattice QCD may play an important role. About half of the uncertainty of $\bar g_0^\theta$ results from the
QCD contribution to the neutron-proton mass difference, the other half from the ratio of the $u$ to
$d$ quark masses. In the last ten years the determination of the latter quantity has improved from a range of 
$0.3$--$0.7$ ({\it cf.} Ref.~\cite{pdg:2004}) 
to $0.38$--$0.58$ ({\it cf.} Ref.~\cite{pdg:2012}) because of improved lattice QCD calculations.
 The standard for lattice calculations is now
``full QCD'' with   two light ($u$ and $d$ sea) quarks plus one heavy ($s$ sea)  quark~\cite{Colangelo:2010et, Aoki:2013ldr}. 
Direct lattice determinations exist already for the strong-interaction contribution to the neutron-proton mass
difference~\cite{Beane:2006fk} which will be improved in the future. In addition, once the electromagnetic contribution to hadronic ground state masses
can be fully included in the simulations (see Ref.~\cite{Borsanyi:2013lga} for the state of the art which, however, does not include all dynamical effects yet), the physical neutron-proton mass difference can be included in the analysis as well which should lead to
improved values of both $m_u/m_d$ and the strong-interaction contribution to the neutron-proton mass difference. This in turn will lead to a
 reduction of the uncertainty of $\bar g_0^\theta$.

 The situation for $\bar g_1^\theta$ is different to the extend
that only half of the uncertainty given in Eq.~(\ref{g1theta}) stems from the LEC $c_1$. This LEC is related
to the nucleon $\sigma$-term and is open for improvements from lattice QCD or from studies of the $\pi N$-system. The other half results from the NDA
estimate of an additional contribution to the isospin-breaking $\slashPT$ pion-nucleon vertex which cannot be
traced back to the $\sigma$-term. 
In summary, future lattice calculations might reduce the uncertainty of  $\bar g_1^\theta$ 
listed in Eq.~(\ref{g1theta}) by a factor of two, such that the two-nucleon contribution to the deuteron EDM,
the analog of Eq.~\eqref{thetatest_d}, will be predicted with only a ($30$--$40$)\% error.

Even the comparably small nuclear uncertainty can be reduced by about a factor of two by 
replacing the $PT$-even N$^2$LO interactions and pertinent wave functions
by their N$^3$LO counterparts (including  N$^3$LO three-body forces 
in the helion and triton cases)~\cite{J.Bsaisou,Jan_new}.
The application of these improved chiral potentials and wave functions together
with the envisaged lattice improvements  might finally reduce the uncertainties of the two-nucleon contributions to the helion and triton EDMs listed in Eq.~(\ref{thetatest_3H}) to a ($20$--$30$)\% error.

The improvements in the chiral potentials and wave functions hold of course also 
for the BSM scenarios. 
The uncertainty of the two-nucleon contribution to the deuteron EDM (Eq.~(\ref{dEDMfull})) 
and to the tri-nucleon EDMs (Eq.~(\ref{tri})) 
might then be reduced by about $50\% $.
Nevertheless the biggest unknowns in these cases are the hadronic inputs, the LECs
$\bar g_0$ and $\bar g_1$  (and, in addition, the strength of the three--pion vertex, ${\bar \Delta}$, in the mLRSM scenario).
The calculation of the \mbox{$\slashPT$} three-point correlators, pion-two-nucleon for $\bar g_{0,1}$ and three-pion for $\bar \Delta$, seems 
to be a task that probably only lattice calculations can address in the future. The same can be said of the single-nucleon EDMs generated by dimension-six sources.

\subsection{Summary}
 In this work we have investigated,  by using four different models
 of flavor-diagonal $\CP$ violation, how  distinct $\CP$ scenarios leave their
footprint in EDMs of different systems.
 Such a study  has to be performed in several steps: First,  the models
 are analyzed at  some high-energy scale  where perturbation theory
applies.  We studied the Standard Model (with massless neutrinos) including the QCD
  $\bar\theta$ term, the minimal left-right symmetric model (mLRSM), an
aligned two-Higgs model (a2HDM) and, briefly, the minimal supersymmetric
extension of the Standard Model (MSSM).
In each case we have investigated the pertinent $\CP$-odd sources and how they induce, at lower energies, effective $\slashPT$ operators of dimension six (in the Standard Model case, the $\tb$ term appears directly and is of dimension four). Because symmetries and field content differ between the four scenarios, they give rise to different (sets of) effective operators. In the Standard Model the only relevant operator is the $\tb$ term, in the mLRSM the dominant operator is a $\slashPT$ four-quark operator with nontrivial chiral and isospin properties, while in the a2HDM scenario the quark EDM, chromo-EDM, and the Weinberg three-gluon operator are all relevant. In the MSSM the situation is, in general, similar to the a2HDM,
although distinctive  scenarios are also possible within the MSSM
framework, \textit{cf}. Sect.~\ref{suse:MSSM}.

The next step involves the evolution of the resulting operators to the low energies where the experiments take place. This can be done perturbatively down to an energy around the chiral-symmetry-breaking scale by use of one-loop QCD renormalization-group equations. To go to even lower energies, nonperturbative techniques are required. In this work we have extended the Lagrangian of chiral perturbation theory ($\chi$PT)  to obtain an EFT describing $\slashPT$ interactions among pions, nucleons, and photons which are the relevant degrees of freedom for hadronic and nuclear EDMs. $\chi$PT allows for a systematic derivation of the operator structure of the $\slashPT$ hadronic interactions. However, only some low energy
parameters induced by the $\tb$ term can be controlled quantitatively. For all other scenarios
each interaction comes with an unknown strength, traditionally called low energy constant (LEC),  whose size cannot be obtained from symmetry arguments alone. Nevertheless, as argued here and elsewhere \cite{BiraEmanuele, Jan_2013, deVries:2012ab}, symmetry considerations still provide important clues on the hierarchy of the various interactions.  

It was demonstrated that
 the $\slashPT$ dimension-four and -six operators appearing in the various scenarios transform differently under chiral and isospin rotations
 which carries over to the parameters of the
  chiral Lagrangian. In the pionic and pion-nucleon sector the most important interactions are given by
\begin{equation}\label{disc1}
\mathcal L=  \bar g_0 \Nb \boldpi\cdot \boldtau N+ \bar g_1 \Nb \pi_3 N-\bar \Delta \frac{\pi_3 \boldpi^2}{2\Fp} \,\,\,,
\end{equation}
where the relative sizes of the three LECs $\bar g_{0,1}$ and  $\bar \Delta$ depend crucially on the $\CP$-odd scenario under investigation. In particular, one finds for the ratio $\bar g_1/\bar g_0$ \cite{Pospelov_piN, BiraEmanuele, Jan_2013, deVries:2012ab}  
\begin{equation}\label{disc2}
\frac{\bar g_1^{\tb}}{ \bar g_0^{\tb}} = -0.2\pm0.1\,\,,\qquad \frac{\bar g_1^{\mathrm{LR}}}{\bar g_0^{\mathrm{LR}}} = -50\pm 25\,\,,\qquad \left|\frac{\bar g_1^{\mathrm{H}}}{\bar g_0^{\mathrm{H}}}\right| \simeq 1\,\,,\qquad \left|\frac{\bar g_1^{\mathrm{MSSM}}}{\bar g_0^{\mathrm{MSSM}}}\right| \simeq 1\,\,\,,
\end{equation}
where the uncertainties are largest in the a2HDM and MSSM scenarios. In addition, the three-pion vertex proportional to $\bar \Delta$ only appears at leading order in the mLRSM \cite{deVries:2012ab}, while it provides a next-to-next-to-leading-order correction in the other scenarios.

Despite these differences in the pionic and pion-nucleon sector, the EDMs of the nucleons do not necessarily show a distinct pattern, since in all scenarios the nucleon EDMs obtain leading-order contributions from the short-range operators of Eq.~\eqref{Ngamma} \cite{Borasoy:2000pq, BiraHockings, Narison:2008jp,ottnad, deVries2010a, Mer11, Guo12}. The sizes of the corresponding LECs are not constrained by chiral symmetry which means that 
our approach has little predictive power in the single-nucleon sector. Model calculations or estimates can provide some information on the sizes of the LECs, but the uncertainties are large. In case of the $\tb$ term, lattice results are available which provide additional information, but, at the moment, the uncertainties are still too large to draw firm conclusions. In summary, measurements of the nucleon EDMs are not enough to disentangle the various scenarios. 

Dedicated storage rings might allow  for measurements of  EDMs of light ions. Because $\chi$PT allows for a unified description of nucleons and (light) nuclei, light-nuclear EDMs can be calculated in terms of the LECs in Eq.~\eqref{disc1} and the nucleon EDMs. The associated nuclear uncertainties can be systematically estimated and turn out to be small compared to the uncertainties in the sizes of the LECs themselves in contrast to calculations of some of the heavier systems. One reason why the storage ring proposals are so interesting is that nuclear EDMs already depend at tree level on the interactions in Eq.~\eqref{disc1}, providing
direct access to the  nontrivial relations of Eq.~\eqref{disc2}, in contrast to the single-nucleon EDMs, where these interactions
contribute only at one-loop level and
are masked by the presence of the additional short-ranged operators mentioned above.  Since different light-nuclear EDMs\footnote{in this work we looked at the deuteron, helion, and triton EDMs, but other light-nuclear EDMs can be calculated in similar fashion.} 
depend on the same set of LECs with different relative weight \cite{deVries2011b}, the dependence on $\bar g_{0,1}$ (and possibly $\bar \Delta$) can be isolated and the hierarchy presented in Eq.~\eqref{disc2} can be studied experimentally, once measurements are performed on the EDMs of different light ions as discussed in Sect.~\ref{sec:test}.

It should be stressed that the models discussed in this paper were chosen to illustrate the potential as well as limitations of
detailed analyses of various EDM measurements. Clearly, this choice is to some extend arbitrary and does by no means exhaust 
the possible options for  physics beyond the Standard Model. However, it should have become clear that the methods applied in
this and earlier works are quite general and can also be used to analyze the signatures of other models for $\CP$ violation beyond the Kobayashi-Maskawa mechanism.

In summary we have argued that an experimental program aimed at measurements of EDMs of light nuclei is very promising. Such measurements have sufficient sensitivity to probe scales where well-motivated scenarios of physics beyond the Standard Model are expected to appear. In addition, we have demonstrated that these measurements are expected to play an essential role in unraveling the origin(s) of $\CP$ violation.

\subsection*{Acknowledgements}
We thank Daniel Boer, Feng-Kun Guo, Susanna Liebig, Emanuele Mereghetti, David Minossi, Andrea Shindler, Rob Timmermans, and  Bira van Kolck for useful discussions.
This work is supported in part by the DFG and the NSFC
through funds provided to the Sino-German CRC 110 ``Symmetries and
the Emergence of Structure in QCD'' (Grant No. 11261130311).
We acknowledge the support of the European Community-Research Infrastructure 
Integrating Activity ``Study of Strongly Interacting Matter'' (acronym
HadronPhysics3, Grant Agreement n. 283286) under the Seventh Framework Programme of EU.

\appendix
\section{The minimal left-right symmetric model}
 \label{LRmodel}
In this appendix we discuss the $\slashPT$ dimension-six operators
arising in the mLRSM that can contribute to Eq.~\eqref{introeq}.
In particular we derive the operator in Eq.~\eqref{Wcurrent} which
gives the dominant contribution to hadronic EDMs. As mentioned in
Sect.~\ref{mLRSM}, this is not the only  {$\slashPT$ operator induced}
 at the electroweak scale. {The (C)EDM  operators of light
   quarks are also generated,  but only through loop diagrams which suppresses the EDMs and CEDMs  $d_q$ and
${\tilde d}_q.$  
 In addition, the light-quark EDMs and CEDMs are proportional to 
 a  small quark mass,}
  which further suppresses their contribution to hadronic EDMs with respect to the four-quark operators arising from Eq.~\eqref{Wcurrent}~\cite{Xu:2009nt,An:2009zh}.
The Weinberg three-gluon operator can be produced as well, but only at
the two-loop level and its contribution to EDMs is
negligible~\cite{Xu:2009nt,An:2009zh}. Finally, $\slashPT$ four-quark
operators are induced by tree-level exchange of the additional,
{non-SM-like Higgs particles of the model that have $\CP$-violating
  couplings to quarks.}  However, the four-quark operators involving
light quarks are suppressed by small Yukawa couplings. 
In combination with the fact that the additional Higgs bosons giving rise to these four-quark operators should
be heavy, with masses exceeding $15\, \text{TeV}$ in order to evade FCNC constraints \cite{Zhang:2007da}, 
 we can neglect such four-quark operators~\cite{Xu:2009nt,An:2009zh}.

Thus, for {hadronic} 
 EDMs the most important {interaction} is the right-handed
 current {interaction}
 in Eq.~\eqref{Wcurrent}, which is produced after integrating out the
 $W_R^\pm$ boson. This operator arises from the interaction between
 the charged gauge-bosons, $W_{L,R}^\pm$, and the bidoublet
   {$\phi$ defined in  Eq.~\eqref{LRphi}.}
 In fact, it is the kinetic term of {the} bidoublet which is
 responsible for the mixing between 
the $W_L^\pm$ and $W_R^\pm$ bosons, which in turn gives rise to the
operator in Eq.~\eqref{Wcurrent}. 
{Using that}
\bea
\mathcal{D}_\mu \phi  = \partial_\mu \phi +i\frac{g_L}{2} W_{L\mu}^a\tau^a \phi -i\frac{g_R}{2} \phi W_{R\mu}^a\tau^a\,\,\,,
\eea
{where $g_{L,R}$ are the coupling constants of the $SU(2)_{L,R}$ gauge groups which are equal in the mLRSM, $g_L=g_R$, the kinetic term of the bidoublet is given by}
\bea
\vL = \text{Tr}[(\mathcal{D}_\mu \phi)^{\dagger}(\mathcal{D}^\mu \phi)]=
 \frac{i g_R}{\sqrt{2}}\text{Tr}\bigg[\bma 0 &
 W_{R\mu}^{+}\\W_{R\mu}^{-} &0\ema \phi^{\dagger}\mathcal{D}^\mu \phi
 \bigg] +\text{h.c.}+\dots\,\,\,. \label{WRterms}
\eea
{Here} we only kept terms bilinear in $W_R^\pm$. We can
now integrate out $W_R^\pm$ to obtain
\bea\vL_{W_R} = \frac{i g_R\sq}{2M_R\sq}\text{Tr}\bigg[\bma 0 & J_{R\mu}^+\\J_{R\mu}^- &0\ema \phi^{\dagger}D^\mu_L \phi \bigg]+\text{h.c.}+\dots\,\,\,,
\label{Wcurrent2}\eea
with $J^-_{R\mu} = \overline U_RV_R\g_\mu D_R$ and $J^+_{R\mu} =
(J^-_{R\mu})^{\dagger}$, {while $V_R$ is the quark mixing matrix of the
  right-handed sector,  and $M_R\approx g_Rv_R$ is the mass of $W_R^\pm$. Moreover,
 \bea
D^\mu_L \phi  = \partial^\mu \phi +i\frac{g_L}{2} W_{L}^{a\mu} \tau^a \phi\,\,\,.
\eea
The form of the interaction in Eq.~\eqref{Wcurrent} is already visible
in Eq.\ \eqref{Wcurrent2}. 
 {It only remains to integrate out the heavy Higgs fields. To do
   so we write the bidoublet $\phi$ in terms of two $SU(2)_L$ doublets,
\bea \phi = (\phi_1 , \phi_2)\,\,,\nn \qquad
\phi_1 \equiv \bma \phi_1^0 \\  \phi_1^-\ema\,\,, \qquad \phi_2 \equiv
\bma \phi_2^+ \\  \phi_2^0\ema\,\,\, .
\eea
Since the field $\varphi$ that corresponds to the SM Higgs field
 is a doublet under $SU(2)_L$ as well, it is taken to be a linear
 combination of these fields. 
 The remaining linear combination then only involves Higgs fields that
 are, by assumption, heavy. To good approximation these Higgs fields are given, in terms of
 the fields in the mass
 basis, by \cite{Zhang:2007da}:}
\bea \label{LRApPMB}
\varphi = \bma -G_L^+ \\ (h^0 + i G_Z^0)/\sqrt{2} \ema\,\,, \qquad \varphi_H = \bma H_2^+ \\ (H_1^0 + i A_1^0)/\sqrt{2} \ema\,\,\,.
\eea
{Here $G_{L}^+$ and $G_Z^0$ are the would-be Goldstone boson
  fields that get absorbed by the $W_L^+$ and $Z_L$ fields,
  respectively, while $h^0$ corresponds to
    the SM Higgs boson. The fields appearing in $\varphi_H$ are 
    assumed to be  heavy.
The basis transformation between the fields $\phi_{1,2}$ and those of Eq.~\eqref{LRApPMB}
  is given by~\cite{Zhang:2007da} }
\bea
\bma \varphi \\ \varphi_H \ema = \frac{1}{\sqrt{1+\xi\sq}}\bma -1 & \xi e^{-i \al} \\ \xi e^{i\al} & 1 \ema \bma \tilde \phi_1 \\ \phi_2\ema\,\,\,,
\label{basistransf}\eea
where $\xi  = \kappa'/\kappa$ and  $\tilde \phi = i \tau_2 \phi^{*}$. 
{With} Eq.\ \eqref{vev} we can check that $\langle \varphi\rangle = \sqrt{\kappa\sq+\kappa^{\prime\,2}} = v/\sqrt{2}$ while $\langle \varphi_H\rangle = 0$. 

Using Eq.~\eqref{basistransf} to {rewrite} Eq.~\eqref{WRterms} in
terms of the {fields in the mass basis} and keeping only terms containing the light fields $\varphi$, we obtain
\bea 
\vL_{W_R} &= &\frac{ig_R\sq}{2M_R\sq}\frac{1}{1+\xi\sq} \text{Tr}\bigg[\bma J^+_{R\mu}\xi e^{-i \al}\varphi^{\dagger}\\ J^-_{R\mu}\tilde \varphi^{\dagger}\ema D^\mu (\tilde \varphi, \,\xi e^{i\al} \varphi )\bigg] +\text{h.c.}+\dots \nn\\
&=&\frac{ig_R\sq}{2M_R\sq}\frac{\xi}{1+\xi\sq} \big[e^{i\al}\tilde\varphi^{\dagger}(D^\mu \varphi)J^-_{R\mu} +e^{-i\al}\varphi^{\dagger}(D^\mu \tilde\varphi)J^+_{R\mu}\big]+\text{h.c.}+\dots\nn\\
&=&\frac{ig_R\sq}{M_R\sq}\frac{\xi}{1+\xi\sq}e^{i\al}\tilde\varphi^{\dagger}(D^\mu \varphi)J^-_{R\mu}+\text{h.c.}+\dots\,\,\,, 
\eea
where we used $\big[i\tilde\varphi^{\dagger} (D_\mu \varphi)\big]^{\dagger} =  i\varphi^{\dagger}(D_\mu\tilde\varphi) $. Finally, a comparison with Eq.~\eqref{Wcurrent} shows that
\bea \Xi_1 = \frac{g_R\sq}{M_R\sq}\frac{\xi}{1+\xi\sq}e^{i\al}V_{R}^{ud} \simeq\frac{1}{\kappa\sq+\kappa^{\prime\,2}}\frac{\kappa \kappa'}{v_R\sq}e^{i\al}V_{R}^{ud}\simeq -\frac{2}{v\sq}\sin\zeta\, V_R^{ud }e^{i\al}\,\,\,.
\eea

\section{The aligned two-Higgs doublet model}
 \label{A2HDMAppendix}
In this appendix we 
  {discuss how the low-energy   ${\slashPT}$ Lagrangian in Eq.~\eqref{2HDM1GeV} comes about in the a2HDM
    with the parameter specifications in Eq.~\eqref{assumptions}.  

\subsection{ $\CP$-violating four-quark operators}

 $\CP$-violating four-quark operators with net flavor number zero are
 induced, in the model with the parameters in Eq.~\eqref{assumptions},
 already at tree level by the exchange of the neutral Higgs bosons $H$
 and $A$ and of charged Higgs bosons $H^\pm$ with Yukawa interactions
 in Eq.~\eqref{yukawa3}.  Because we assume 
  $H$ and $A$ to be (nearly) mass-degenerate, $M_H\simeq M_A \simeq
  M$, the relation Eq.~\eqref{simplification} can be applied to the computation of the
   coefficients of these operators. Then, as already briefly mentioned
  in Sect.~\ref{2HDM}, the exchange of $H$ and $A$  induces operators of the type
   $(\bar{u}u) (\bar{d}i\gamma_5 d)$ and  $(\bar{d}d)
   (\bar{u}i\gamma_5 u)$ 
  with coefficients $\pm m_u m_d {\rm
    Im}(\varsigma_{u}^*\varsigma_{d})/(v^2M^2)$, where $u$ $(d)$
   denotes here any of the up-type (down-type) quarks.
  The operators that involve light quarks only
  are severely suppressed by the factor $m_u m_{d} /v^2$. 
The  contribution of these operators to the EDM of a nucleon turns out
    to be, after the assumptions in Eq.~\eqref{assumptions}, significantly smaller
 than  the two-loop dipole contributions discussed below in  
  the appendices~\ref{suse:apEDM} and~\ref{suse:apW}.

The tree-level exchange of the
 charged Higgs bosons $H^\pm$ between quarks,
 with Yukawa interactions given in Eq.~\eqref{yukawa3}, induces at tree-level
 the ${\slashPT}$ operators  $(\bar{u}d)(\bar{d}i\gamma_5 u)$ and
 $(\bar{u}i\gamma_5 d)(\bar{d}u)$ with coefficients $2 m_u m_d |V_{ud}|^2{\rm
    Im}(\varsigma_{u}^*\varsigma_{d})/(v^2M_+^2)$. The above statements on
  the size of the four-quark contributions induced by neutral Higgs
  boson exchanges apply also here. 

The operators containing heavy quarks can (partially) overcome these
suppression factors. However, these operators do not contribute directly to nucleon EDMs.
 Operators with two heavy quark fields can, after integrating out the
 heavy quarks,  induce dimension-seven operators of the form
 $(\bar q q) \epsilon^{\alpha\beta\mu\nu} G^a_{\alpha\beta} G^a_{
   \mu\nu}$ 
and $(\bar q i \g_5q) G^a_{\mu\nu} G^{a
  \,\mu\nu}$~\cite{Anselm:1985cf,Demir:2003js}, where $q$ denotes a
light quark.
The size of the contributions of these operators to the nucleon EDM
has been estimated in Ref.~\cite{Demir:2003js} and also
turns out to be significantly smaller than the contributions coming
from the two-loop dipole diagrams to be discussed below.

These considerations justify that we neglect the contributions of
four-quark operators to the low-energy effective Lagrangian in Eq.~\eqref{2HDM1GeV}
 in the a2HDM model with  the parameter specifications in Eq.~\eqref{assumptions}.

The exchange of the neutral Higgs bosons $H$
 and $A$ between quarks and leptons $\ell$
 induces  $\CP$-violating semileptonic four-fermion operators
 $(\bar{q}q) (\bar{\ell}i\gamma_5 \ell)$ (and $q\leftrightarrow \ell$)
 with coefficients  $\pm m_q m_\ell {\rm
   Im}(\varsigma_{q}^*\varsigma_{\ell})/(v^2M^2)$.
  These are of potential importance for $T$-violating effects in
  paramagnetic atoms (\textit{cf}. Sect.~\ref{sec:eEDM}).
 However, if
\begin{equation} \label{ap:relqlp}
 {\rm Im}(\varsigma_{q}^*\varsigma_{\ell}) = {\cal O}\left( {\rm
     Im}(\varsigma_{u}^*\varsigma_{d}) \right) \, ,
\end{equation}
then the electron EDM induced by two-loop Barr-Zee diagrams
\cite{Barr:1990vd} dominates by far the contribution
 to the $T$-violating effect 
 in  the ThO molecule that was recently searched for in Ref.~\cite{Baron:2013eja}.

\subsection{Contributions to the quark EDMs and chromo-EDMs}
\label{suse:apEDM}

 In general, $\CP$-violating flavor-diagonal neutral Higgs boson
 exchanges induce quark (C)EDMs already at one-loop. Because these
 one-loop terms scale with the third power of the quark mass (modulo logs),
 $d_q^{(1l)}, {\tilde d}_q^{(1l)} \sim m_q^3/(v^2M^2)$, they are, in
 the case of light quarks, suppressed as compared to
 the two-loop Barr-Zee contributions. Although these are nominally
 suppressed by an additional loop factor $\alpha/(4\pi)$, respectively
 $\alpha_s/(4\pi)$, where $\alpha$ $(\alpha_s)$ is the QED (QCD)
 coupling, they involve only one power (modulo logs) of $m_q$.
 In the a2HDM with the specifications of Eq.~\eqref{assumptions}, the one-loop
 exchanges of the neutral Higgs bosons $H$ and $A$ cannot, in fact,
 generate a one-loop contribution to a quark (C)EDM. This follows
 from Eq.~\eqref{simplification}. The  exchange of a
 charged Higgs boson does generate a one-loop contribution.
 For instance, the EDM of the $d$ quark receives a contribution
 $d_q^{(1l)}(H^+) \sim  2 m_d m_u^2 |V_{ud}|^2   {\rm
     Im}(\varsigma_{u}^*\varsigma_{d})/(v^2M_+^2)$.
 But also these one-loop terms are subdominant compared to the 
 two-loop terms that we now discuss and can therefore be neglected.

 For the general a2HDM with the Yukawa interactions of Eq.~\eqref{yukawa3} and the couplings in Eqs.~\eqref{eq:y0u2hdm} and \eqref{eq:y0d2hdm}
 the Barr-Zee-type diagrams involving a $\CP$-violating 
 neutral spin-zero particle and a quark in the loop induce the 
 following contribution to the quark EDM and CEDM, respectively \cite{Barr:1990vd, Gunion:1990iv,Jung:2013hka,Abe:2013qla}:}
\bea
{d_q(\mu_H;\varphi^0,\,q)}& =& 24e Q_q m_q\frac{\alpha}{(4\pi)^3 v\sq}\sum_{q',i} Q_{q'}\sq\bigg[ f\left(\frac{m_{q'}\sq}{M_i\sq}\right)\text{Re}\, y^i_q \text{Im}\, y^i_{q'}+g\left(\frac{m_{q'}\sq}{M_i\sq}\right)\text{Re} \,y^i_{q'} \text{Im}\, y^i_q\bigg]\,\,\,,\nn\\
{\tilde d_q(\mu_H;\varphi^0,\,q)} &=& -4m_q \frac{g_s\alpha_s}{(4\pi)^3 v\sq}\sum_{q',i}\bigg[ f\left(\frac{m_{q'}\sq}{M_i\sq}\right)\text{Re}\, y^i_q \text{Im}\, y^i_{q'}+g\left(\frac{m_{q'}\sq}{M_i\sq}\right)\text{Re}\, y^i_{q'} \text{Im}\, y^i_q\bigg]\,\,\,,
\label{BZq}
\eea
where $e>0$, $Q_u=2/3, Q_d=-1/3$, and the label $\mu_H$ indicates that these are the quark (C)EDMs
  at a scale $\mu_H \sim m_t \sim M_\varphi$. The  QCD coupling is understood to be evaluated at the scale $\mu_H$.
 For a neutral spin-zero particle and a $W^\pm$ boson in the loop
  one gets \cite{Barr:1990vd, Gunion:1990iv,Jung:2013hka,Abe:2013qla}:}
\bea
{d_q(\mu_H;\varphi^0,\,W^\pm)} = -4e Q_q m_q\frac{\alpha}{(4\pi)^3 v\sq}\sum_{i} \bigg[ 3f\left(\frac{M_W\sq}{M_i\sq}\right)+5 g\left(\frac{M_W\sq}{M_i\sq}\right)\bigg]\text{Im}\, \left(y^i_q R_{i1}\right)\,\,\,,
\label{BZw}
\eea 
where 
\bea
f(z) \equiv \frac{z}{2}\int_0^1 dx \frac{1-2x(1-x)}{x(1-x)-z}\ln \frac{x(1-x)}{z}\,\,, \qquad g(z) \equiv \frac{z}{2}\int_0^1 dx \frac{1}{x(1-x)-z}\ln \frac{x(1-x)}{z}\,\,\,.
\eea

We now apply the specifications of Eq.~\eqref{assumptions}.
 Then the contribution in Eq.~\eqref{BZw} is zero, because $R_{11}=1$, $R_{21}=R_{23}=0$ and $y^1_q=1$.
 Eqs.~\eqref{assumptions} and \eqref{simplification} imply that  up-type quarks in the fermion loop
  contribute only to the (C)EDM of a down-type quark and vice versa. Therefore, diagrams with a top quark in the loop
  contribute only to the (C)EDM of the $d$ quark. Diagrams with quarks $q\neq t$ in the loop
   are suppressed by at least roughly two orders of magnitude as compared to the $t$-quark contribution, because
  of smaller Yukawa couplings. (This is reflected in the significantly smaller magnitudes of the respective values
   of the functions $f$ and $g$.) Therefore, in the a2HDM with the assumptions in Eq.~\eqref{assumptions}, the 
   EDM  and CEDM of the  $u$ quark in the low-energy effective Lagrangian can be neglected
     as compared to those of the $d$ quark.
  There are also contributions to $d_q$ from charged leptons in the fermion loop but, assuming that the relation in Eq.~\eqref{ap:relqlp}
   holds, these  can also be neglected as compared to the $t$-quark contribution to $d_d$.

{ Another set of
  Barr-Zee type contributions to $d_q$  involves charged Higgs boson exchange \cite{BowserChao:1997bb,Jung:2013hka,Abe:2013qla}.
 They contribute significantly to the $d$-quark EDM only, while the $u$-quark EDM is again negligible~\cite{Jung:2013hka,Abe:2013qla}:
\bea
{d_d(\mu_H; H^\pm)} = m_d\frac{12M_W\sq}{(4\pi v)^4}|V_{tb}|\sq|V_{ud}|\sq \text{Im}\,(\varsigma_u^*\varsigma_d)(eQ_t F_t+e Q_b F_b)\label{ChargedqEDM}\,\,\,,
 \eea
where
\bea
F_q = \frac{T_q(z_{H^\pm})-T_q(z_W)}{z_{H^\pm}-z_W}\,\,,\qquad z_i\equiv \frac{M_i\sq}{m_t\sq}\,\,\,,\eea
and
\bea
T_t(z) &=& \frac{1-3z}{z\sq}\frac{\pi\sq}{6} + \left(\frac{1}{z}-\frac{5}{2}\right)\ln z-\frac{1}{z}-\left(2-\frac{1}{z}\right)\left(1-\frac{1}{z}\right)\text{Li}_2(1-z)\,\,\,,\nn\\
T_b(z) &=& \frac{2z-1}{z\sq}\frac{\pi\sq}{6} + \left(\frac{3}{2}-\frac{1}{z}\right)\ln z +\frac{1}{z}-\frac{1}{z}\left(2-\frac{1}{z}\right)\text{Li}_2(1-z)\,\,\,.
\eea
{
This contribution to $d_d$ is not affected by the parameter choices in Eq.~\eqref{assumptions}.

Additionally, there are contributions to the quark EDMs through diagrams which are similar to those that gave rise to Eq.~\eqref{ChargedqEDM}, but where the virtual quark loop is replaced by a loop involving the spin-zero fields \cite{Abe:2013qla,Inoue:2014nva}. These diagrams are proportional to a different $\CP$-violating parameter than the one encountered so far. 
 Although these diagrams generate $u$- and $d$-quark EDMs of similar size,
  the contributions from these diagrams are smaller by a factor of a few than the ones in Eqs.~\eqref{BZq}, see Ref.~\cite{Abe:2013qla}. 
 In addition, they do not contribute to the quark CEDMs which play the dominant role in  our analysis. 
 Thus, under the assumption that the $\CP$-violation parameters are of similar 
 magnitude, these diagrams are expected to be less important than the CEDMs. Therefore, we neglect them in our analysis.

In summary, we obtain in the a2HDM with the parameter specifications in Eq.~\eqref{assumptions} that at a high scale $\mu_H$ 
  the $d$-quark (C)EDM is significantly larger than
 the corresponding dipole moment of the $u$ quark. The $d$-quark EDM and CEDM are  given by, putting $\mu_H=m_t$:}
\bea
d_d(m_t)& =&  e \frac{Q_dm_d\alpha}{(4\pi)^3 v\sq}\text{Im}\,(\varsigma_u^*\varsigma_d)\bigg(\frac{32}{3}\bigg[ 
f\left(\frac{m_{t}\sq}{M\sq}\right)+g\left(\frac{m_{t}\sq}{M\sq}\right)\bigg]+\frac{3}{s_w\sq}|V_{ud}|\sq|V_{tb}|\sq\big[F_b-2F_t\big]\bigg),\,\,\,\,\label{2HDMemt}\\
\tilde d_d(m_t) &=& - 4 m_d \frac{g_s \alpha_s}{(4\pi)^3 v\sq}\text{Im}\,(\varsigma_u^*\varsigma_d)\bigg[ f\left(\frac{m_{t}\sq}{M\sq}\right)+g\left(\frac{m_{t}\sq}{M\sq}\right)\bigg]\,\,\, .
\label{2HDMCmt}
\eea
{They depend on a common unknown factor 
 $\text{Im}\, (\varsigma_u^*\varsigma_d)$ that signifies non-KM $\CP$ violation of the model.}
 {By renormalization-group evolution down to the scale $\mu=\Lambda_\chi$ we obtain the $d$-quark (C)EDM given in Eqs.~\eqref{2HDM1GeV} and \eqref{2hdm:hatd}.

\subsection{Contributions to the Weinberg operator}
\label{suse:apW}
The leading-order contributions to the Weinberg  operator corresponds to diagrams of the type shown in Fig.~\ref{diagrams}(c). 
  From diagrams that involve $\CP$-violating flavor-diagonal neutral Higgs boson exchange the coefficient of
  the Weinberg operator, i.e., the CEDM of the gluon,  receives, in the 2HDM,  the following contribution~\cite{Weinberg:1989dx,Dicus:1989va,Jung:2013hka}:}
\bea
\label{WeinbergN}
d_W(m_t; \varphi^0)=-\frac{4 g_s^3}{(4\pi)^4v\sq}\sum_{q,i}\, \text{Re}\, y_q^i \text{Im}\, y_q^i \,h(m_q,\,M_i)\,\,\,,\eea
where
\bea
h(m,M) = \frac{m^4}{4}\int_0^1 dx\int_0^1 du \frac{u^3x^3(1-x)}{[m\sq x(1-ux)+M\sq(1-u)(1-x)]\sq}\,\,\,.
\eea 
As we restrict ourselves to the parameters of
  Eq.~\eqref{assumptions}, this contribution will be proportional to $\text{Im}( \varsigma_q^* \varsigma_{q})$ and therefore vanishes.

The exchange of a charged Higgs boson with Yukawa couplings given in  Eq.~\eqref{2HDM1GeV}  leads to
diagrams similar to Fig.~\ref{diagrams}(c).
 In this case both a bottom and top quark are present in the fermion loop.
 The amplitude involves  two different scales,  $M_{+}\sim m_t$, and $m_b$. One may evaluate it in the framework of effective field theory\footnote{
 Alternatively, this two-loop amplitude was computed directly in Ref.~\cite{Dicus:1989va}. This result can then by renormalization-group
 evolution be evaluated at a low scale.}
 \cite{BraatenPRL,Boyd:1990bx,Jung:2013hka}.
 One can  first integrate out the charged Higgs boson and the top quark. This generates a one-loop contribution to 
the  bottom quark CEDM. 
 At the bottom quark threshold this $b$-quark CEDM then induces a one-loop contribution \cite{BraatenPRL} to  $d_W$.
 The first step gives \cite{Boyd:1990bx, Dicus:1989va, Jung:2013hka}:
{\bea \label{Dwbmt}
\tilde d_b(m_t; H^\pm) = -\frac{g_s(m_t)}{8\pi\sq v\sq}m_b(m_t)|V_{tb}|\sq \text{Im} \,(\varsigma_d \varsigma_u^*)\bigg[ x_t \bigg(\frac{\ln x_t}{(x_t-1)^3}+\frac{x_t-3}{2(x_t-1)\sq}\bigg)\bigg]\,\,\,,
\eea
where $x_t = m_t\sq/M_{+}\sq$
 and  $ {m}_b(m_t)$ is the   $\overline{\rm MS}$ mass of the $b$ quark at the scale 
 $\mu=m_t$.  At $\mu = m_b$ this induces a contribution to the Weinberg operator \cite{BraatenPRL},
\bea
d_W(m_b; H^\pm) = -\frac{g_s\sq(m_b)}{32\pi\sq m_b(m_b)}\tilde d_b(m_b; H^\pm)\,\,\,,
\eea
where $m_b(m_b)$ denotes  the $\overline{\rm MS}$ mass at $\mu=m_b$ and
$\tilde d_b(m_b; H^\pm)$ is related to $\tilde d_b(m_t; H^\pm)$ by a renormalization-group
   factor: $\tilde d_b(m_b; H^\pm) =\eta_W^{\prime} \tilde d_b(m_t; H^\pm)$ where we introduced the parameter $\eta_W^{\prime} = \big(\frac{\al_s(m_t)}{\al_s(m_b)}\big)^{-19/46} \simeq 1.3$ \cite{Wilczek:1976ry,BraatenPRL, Degrassi:2005zd, Hisano3, Dekens:2013zca}. 

In summary, we obtain in the a2HDM with the parameter set of Eq.~\eqref{assumptions} the following CEDM $d_W$ of the gluon at the scale $\mu=m_b$:
 \bea \label{2HDMWmb}
d_W(m_b) &=& \eta_W \frac{g_s\al_s}{(4\pi)^3 v\sq}|V_{tb}|\sq \text{Im} \,( \varsigma_u^* \varsigma_d)\bigg[ x_t \bigg(\frac{\ln x_t}{(x_t-1)^3}+\frac{x_t-3}{2(x_t-1)\sq}\bigg)\bigg]\,\,\,,
\eea
where the factors of $g_s$ and $\alpha_s$ are to be evaluated at the scale $\mu=m_b$ while the parameter $\eta_W = \frac{g_s(m_t)}{g_s(m_b)}\frac{m_b(m_t)}{m_b(m_b)}\eta_W' = \big(\frac{\al_s(m_t)}{\al_s(m_b)}\big)^{14/23} \simeq 0.67$ is the resulting
  renormalization-group  factor due to the evolution from the scale $m_t$ to $m_b$. 
 The  renormalization-group evolution of  Eq.~\eqref{2HDMWmb} to the scale $\mu=\Lambda_\chi$ then yields $d_W(\Lambda_\chi)$ given in
  Eqs.~\eqref{2HDM1GeV} and \eqref{2hdm:hatd}.
}

 \newpage

\bibliographystyle{h-physrev3} 
\bibliography{bibliography}

\end{document}